\documentclass[10pt,draftclsnofoot,onecolumn,letterpaper]{IEEEtran} % New style

\usepackage[rgb, x11names]{xcolor} % x11 colors
\usepackage{amsfonts, amsmath, amssymb, bm} % Math; amsthm conflicts with others
\usepackage{hyperref} % Hyper link
\usepackage{paralist} % inparaenum enviroment
\usepackage{breqn} % dmath enviroment
\usepackage{marginnote} % Enable \margin­note for annotation; Used in \comment
\usepackage{todonotes} % Add todos; Option of [disable] will disable all todos
\usepackage{booktabs} % Provides commands, like \toprule, to create high quality ta­bles
\usepackage{array} % Enable fixed width table cell, i.e, p{1.1\iwidth}<{\centering}
\usepackage{graphicx} % Enable to use \includegraphics
\usepackage{marvosym} % Enable \maketitle; Must be included when using llncs
\usepackage{bookmark}
\usepackage[tight]{subfigure}
\usepackage{multirow}
\usepackage{hhline}
\usepackage{cite}
\usepackage{tikz}
\usepackage{pgfplotstable}
\usepackage{pgfplots}
\usepackage{tikz-3dplot}
\usetikzlibrary{mindmap, trees}
\pgfdeclarelayer{background}
\pgfdeclarelayer{foreground}
\pgfsetlayers{background,main,foreground}
\pgfplotsset{compat=1.3,every axis/.append style={font=\scriptsize}, every legend/.append style={font=\scriptsize}}
\tikzstyle{mylabel} = [text=orange, ultra thick, inner sep=1pt, minimum size=15pt, yshift=-6pt, xshift=30pt]
\tikzstyle{fancylabel} = [rounded corners, fill=Blue4, draw=black, very thick, text=white, inner sep=0pt, minimum size=15pt, yshift=0pt]
\tikzstyle{labelwithbackground} = [text=red, fill=none, ultra thick, inner sep=1pt, minimum size=12pt, yshift=-9pt, xshift=9pt]

\newlength{\defbaselineskip}
\newcommand\blfootnote[1] % Used to show acknowledgement when using llncs
{
	\begingroup
	
	\footnotetext{#1}
	\addtocounter{footnote}{-1}
	\endgroup
}
\usepackage{environ}
\label{packages}

\usepackage{graphicx}

\usepackage{import}
\usepackage{color}
\usepackage{pgf-umlsd}
\usepackage{ifthen}
\usepackage{standalone}
\usepackage{tikzinput}
\usepackage{booktabs,tabularx}
\definecolor{bblue}{HTML}{4F81BD}
\definecolor{rred}{HTML}{C0504D}
\definecolor{ggreen}{HTML}{9BBB59}
\definecolor{ppurple}{HTML}{9F4C7C}
\definecolor{oorange}{HTML}{FDAE61}

\begin{document}

\title{
	Multifold Acceleration of Diffusion MRI via Slice-Interleaved Diffusion Encoding (SIDE)
%	Multifold Acceleration of Diffusion MRI via Inter-Volume Incoherent Slice-Undersampling
	}

\author{
    Yoonmi Hong,
    Wei-Tang Chang,
    Geng Chen,
    Ye Wu,
    Weili Lin,
    Dinggang Shen,
    and
	  Pew-Thian Yap
%	  \thanks{\Copyright}
	  \thanks{\RADBRIC}\thanks{\Funding}\thanks{\CoFirstAuthors}
%	\thanks{\CorrAuthor}
	}

\def\RADBRIC{Y.~Hong, W.-T. Chang, G.~Chen, Y.~Wu, W.~Lin, D.~Shen, and P.-T.~Yap are with the Department of Radiology and Biomedical Research Imaging Center (BRIC), University of North Carolina at Chapel Hill, NC, U.S.A. D.~Shen is also with the Department of Brain and Cognitive Engineering, Korea University, Seoul, Korea. Emails: dgshen@med.unc.edu; ptyap@med.unc.edu.}

\def\CorrAuthor{$^{*}$ Corresponding authors.}
\def\CoFirstAuthors{Y.~Hong and W.-T. Chang contributed equally to this work.}

\def\Funding{This work was supported in part by NIH grants (NS093842 and EB006733).}

\maketitle
%\markboth{IEEE Transactions on Medical Imaging}{}

\begin{abstract}
Diffusion MRI (dMRI) is a unique imaging technique for in vivo characterization of tissue microstructure and white matter pathways. 
However, its relatively long acquisition time implies greater motion artifacts when imaging, for example, infants and Parkinson's disease patients.
To accelerate dMRI acquisition, we propose in this paper (i) a diffusion encoding scheme, called Slice-Interleaved Diffusion Encoding (SIDE), that interleaves each diffusion-weighted (DW) image volume with slices that are encoded with different diffusion gradients, essentially allowing the slice-undersampling of image volume associated with each diffusion gradient to significantly reduce acquisition time, and (ii) a method based on deep learning for effective reconstruction of DW images from the highly slice-undersampled data.
%a super-resolution (SR) reconstruction method for slice-undersampled data. 
%Instead of acquiring all slices of diffusion-weighted (DW) image volumes, only a subsample of interleaved slices are acquired. 
%This gives an acceleration factor that is proportional to the subsampling factor. 
%Each DW volume is subsampled with a different slice offset so that complementary information from different DW volumes can be harnessed using graph convolutional neural networks for reconstructing the full DW volumes. 
%In actual acquisition, the slices acquired in a volume in a repetition time (TR) are associated with different slice directions.
%We demonstrate the effectiveness of our method SR reconstruction outperforms competing methods and mitigates partial volume effects. 
Evaluation based on the Human Connectome Project (HCP) dataset indicates that our method can achieve a high acceleration factor of up to 6 with minimal information loss. Evaluation using dMRI data acquired with SIDE acquisition demonstrates that it is possible to accelerate the acquisition by as much as 50 folds when combined with multi-band imaging.
\end{abstract}


\section{Introduction}\label{sec:Introduction}
\IEEEPARstart{D}{iffusion} MRI (dMRI) is widely employed for studying brain tissue microstructure and white matter pathways~\cite{behrens2009diffusion}. Probing water molecules in a sufficient number of diffusion scales and directions is needed for more specific quantification of tissue microenvironments and for more accurate estimation of axonal orientations for tractography.
%precise quantification of tissue microstructure and 
%The conventional diffusion tensor imaging (DTI) is inadequate for accurate representation the complex architecture of crossing white matter fibers. Recently, high angular resolution diffusion imaging (HARDI) techniques have been developed to detect crossing white matter fiber bundles, 
%including Q-ball Imaging (QBI) \cite{tournier2019diffusion}, spherical deconvolution (SD) \cite{tournier2004direct} and Diffusion Spectrum Imaging (DSI) \cite{wedeen2005mapping}. 
This necessitates the acquisition of a large number of diffusion-weighted images and therefore increases acquisition time and susceptibility to motion artifacts, limiting the utility of dMRI for example to pediatric populations.

%A drawback of these techniques is their requirement for a greater number of diffusion encoded acquisitions compared to DTI, leading to an increase in acquisition time \cite{ning2015sparse,yap2016multi,ye2016estimation}. 
%for sufficient coverage of the diffusion wavevector space (i.e., $q$-space). 

%-- What are the limitations of previous SR methods that lead to the proposed method? --
A number of approaches have been proposed to reduce the acquisition time by undersampling  either in $q$-space or in both $k$-space and $q$-space. Ning et al.~\cite{ning2016joint} proposed to subsample in $q$-space incoherently with overlapping thick slices for reconstruction of high-resolution diffusion images. 
%\todo{please check if the highlighted description is correct}
%\rev{To have the incoherent properties of $q$-space subsampling,} 
%high-resolution diffusion parameters are estimated from a set of low resolution diffusion-weighted (DW) images with different slice orientations and diffusion gradient directions, where low resolution images were acquired from different sets of gradients in $q$-space \cite{van2016super}. 
In \cite{chen2018angular}, neighborhood matching in $x$-$q$ space is proposed for angular upsampling of data undersampled in $q$-space. 
Alternatively, high resolution dMRI data can be reconstructed from data undersampled in $k$-$q$ space \cite{cheng2015joint,mani2015acceleration,wu2019diffusion}. 
Cheng et al. \cite{cheng2015joint} proposed a 6-dimensional method for compressed sensing reconstruction of DW images and Ensemble Average Propagators (EAPs) from data subsampled in $k$-$q$ space. 
Mani et al.~\cite{mani2015acceleration} proposed an incoherent $k$-$q$ undersampling scheme, where each point in $q$-space is sampled at different $k$-space locations via random 
%skipping 
interleaves of multi-shot variable density spiral trajectory in $k$-space. Wu et al. \cite{wu2019diffusion} applied different $k$-space sampling patterns by shifting the sampling trajectory in the phase-encoding direction and slice select direction for neighboring points in $q$-space. 

Reconstruction algorithms are typically designed for recovering the lost information in the undersampled data to reconstruct the DW images. Most algorithms recover lost information by assuming some kind of data regularity, such as smoothness \cite{chen2018angular,wu2019diffusion}, sparsity \cite{ning2016joint,cheng2015joint,mani2015acceleration}, and low-rank \cite{shi2016super}.
%Super-resolution (SR) reconstruction algorithms can be used to synthesize DW images from the undersampled data.
%fully-sampled or high-resolution (HR) images from undersampled or low-resolution (LR) images. 
%For example, SR performed the reconstruction using subvoxel-shifted scans in the in-plane \cite{peled2001superresolution} and through-plane \cite{greenspan2002mri} directions.
%Scherrer et al.~\cite{scherrer2012super} proposed a super-resolution (SR) method to reconstruct each isotropic-resolution DW volumes from multiple anisotropic orthogonal DW volumes. 
%This was extended in \cite{scherrer2015accelerated} to estimate the SR tissue model using diffusion compartment imaging. 
%To improve the robustness to noise, the 4D low-rank and total variation regularizations were incorporated into SR reconstructions \cite{shi2016super}.
%Conventionally, the skipped q-space points were reconstructed using (bicubic?), which synthesized the unsampled q-space points from the local proximity of acquired q-space points 
%\todo{is it? -> I think we don't need to explicitly explain bicubic interpolation here. Interpolation is widely used technique for any image resampling problem. Bicubic is one of interpolation methods.}\todo{(citation)}. 
Information recovery can also be achieved using deep learning, which typically learns a non-linear mapping from undersampled to fully-sampled data with the help of training image pairs. This in essence replaces handcrafted image assumptions with characteristics learned directly from the data.
%In contrast to the conventional approach, the SR approach utilized global set of data points for reconstruction rather than local set alone \todo{is it?}. A variety of SR-based methods have been proposed. 
%by solving regularization problem. Recently, deep learning based SR learns a non-linear mapping from LR images to HR images from a large amount of training set.  
%For example, the SR approach was employed to reconstruct diffusion tensor images using random forest, which \todo{explain what random forest does and the achievement unless random forest is a well-known method in your field} \cite{alexander2017image}. 
An example is the image quality transfer framework for reconstructing high-resolution images from their low-resolution counterparts using 
%One of the SR approaches for DTI reconstruction was implemented by combing 
3D convolutional neural networks (CNNs)
% with uncertainty estimation 
\cite{tanno2017bayesian}. 
%SR reconstruction based on a generative adversarial network is proposed in \cite{albay2018diffusion}.
%Albay et al.~\cite{albay2018diffusion} proposed a 
%Generative adversarial networks have also been shown to be effective for super-resolution (SR) reconstruction of high-resolution DW slices from their high-resolution counterparts \cite{albay2018diffusion}. 

%Despite the success of the proposed methods, the performance of SR reconstruction could be further improved by taking into account the correlations between diffusion wavevectors in $q$-space.

%-- Propose SIDE approach and explain how SIDE will outperform the previous SR methods? --
In this paper, we propose to accelerate dMRI acquisition via Slice-Interleaved Diffusion Encoding (SIDE), where each DW image volume is interleaved with slices encoded with multiple diffusion gradients. This is in contrast to conventional encoding schemes that typically encode each volume with a single diffusion gradient. 
SIDE can be seen as slice-undersampling with multiple diffusion gradients.
SIDE differs from existing acceleration techniques in that it does not seek to reduce the repetition time (TR) of a pulsed gradient spin echo (PGSE) echo planar imaging (EPI) experiment. It is designed to acquire more incoherent information without reducing the TR. Reducing the TR can have undesirable consequences such as lower SNR and increased spin-history artifacts.

Reconstructing the full DW volumes from the SIDE DW volumes can be done by first reorganizing the slices in the SIDE volumes to slice stacks according to the diffusion gradients. That is, each slice stack corresponds to a single diffusion gradient. The full DW volumes can then be recovered from the slice stacks with the help of regularity assumptions or image characteristics learned from the data. 
%Such a reconstruction from the subsampled to fully-sampled DW images can be formulated as high-dimensional image synthesis problem.

In this paper, we propose a reconstruction framework that is based on a graph convolutional neural network (GCNN) \cite{bruna2013spectral,henaff2015deep}, which extends CNNs to non-Cartesian domains represented by graphs.
%Conventional 3D based CNN may be applied to our SR reconstruction by considering only spatial neighborhood information.
In dMRI, a 3D image volume is acquired for each point in the diffusion wavevector space, i.e., $q$-space. While the voxels in each image volume reside on a uniform Cartesian grid, i.e., $x$-space, the points in $q$-space might not be distributed in a Cartesian manner. For example, it is common that sampling points are distributed on spherical shells. The GCNN uses a graph to capture the spatial relationships of points in both $x$-space and $q$-space so that the smoothness of the signal in the joint space can be used for effective reconstruction.

%The relationship between the neighboring points in both spatial and wavevector domains can be fully exploited in the form of a graph Laplacian, \rev{where convolution is based on Laplacian in GCNN}.
% need to add GAN, spatial attention

To improve the perceptual quality of DW image, the GCNN is used as the generator in a generative adversarial network (GAN) \cite{goodfellow2014generative}, which have demonstrated impressive results in natural image generation \cite{radford2015unsupervised},\cite{denton2015deep} and in a variety of applications \cite{isola2017image,zhu2017unpaired}. 
%GANs have also been applied to medical image processing, e.g. image synthesis from MRI to CT \cite{nie2018medical} and noise reduction for low-dose CT \cite{wolterink2017generative}. 
The key contributor to the success of GANs is the use of an adversarial loss that forces the generated images to be indistinguishable from real images \cite{zhu2017unpaired}. This is implemented using a discriminator that learns a trainable loss function.

%Squeeze-and-excitation (SE) network is introduced for adaptive recalibration of channel-wise feature responses for image classification \cite{hu2018squeeze}. SE module can be combined with any state-of-the-art architectures at minimal computational cost and provides significant performance improvement. In \cite{roy2018concurrent}, Roy et al. proposed three variants of SE module and showed consistent improvement of performance in image segmentation. We assume that spatial information is more informative in our problem. Hence, we adopt spatial SE (sSE) module, which squeezes along channels and excites or reweights spatially.

Part of this paper has been reported in previous conference publications \cite{hong2019multifold,hong2019reconstructing}. Herein, we provide a more detailed description of our method and extensive experimental results that are not part of the conference paper. % MICCAI paper will be added
The rest of the paper is organized as follows. In Section~\ref{sec:methods}, we describe the SIDE acquisition scheme and the details of the GCNN-based reconstruction framework. In Section~\ref{sec:experiment}, we demonstrate the effectiveness of our method with extensive experiments using retrospectively generated using the data from the Human Connectome Project (HCP) and in vivo human brain data acquired using SIDE. 
Finally, we discuss limitations and future directions in Section~\ref{sec:discuss} and conclude in Section~\ref{sec:conclusion}.

\section{Methods}\label{sec:methods}
In this section, we first describe the SIDE acquisition strategy. Fully-sampled DW images are reconstructed from the SIDE volumes by learning a non-linear mapping based on a GCNN. The graph convolutional operation in our GCNN is based on fast localized spectral filtering \cite{defferrard2016convolutional}. Note that undersampling can be applied to arbitrary scan direction (e.g., axial, coronal, and sagittal). In this paper, we focus on the commonly used axial acquisition.

\subsection{Slice-Interleaved Diffusion Encoding (SIDE)}
\begin{figure*}[t]
	\centering
	\includegraphics[width=0.75\textwidth]{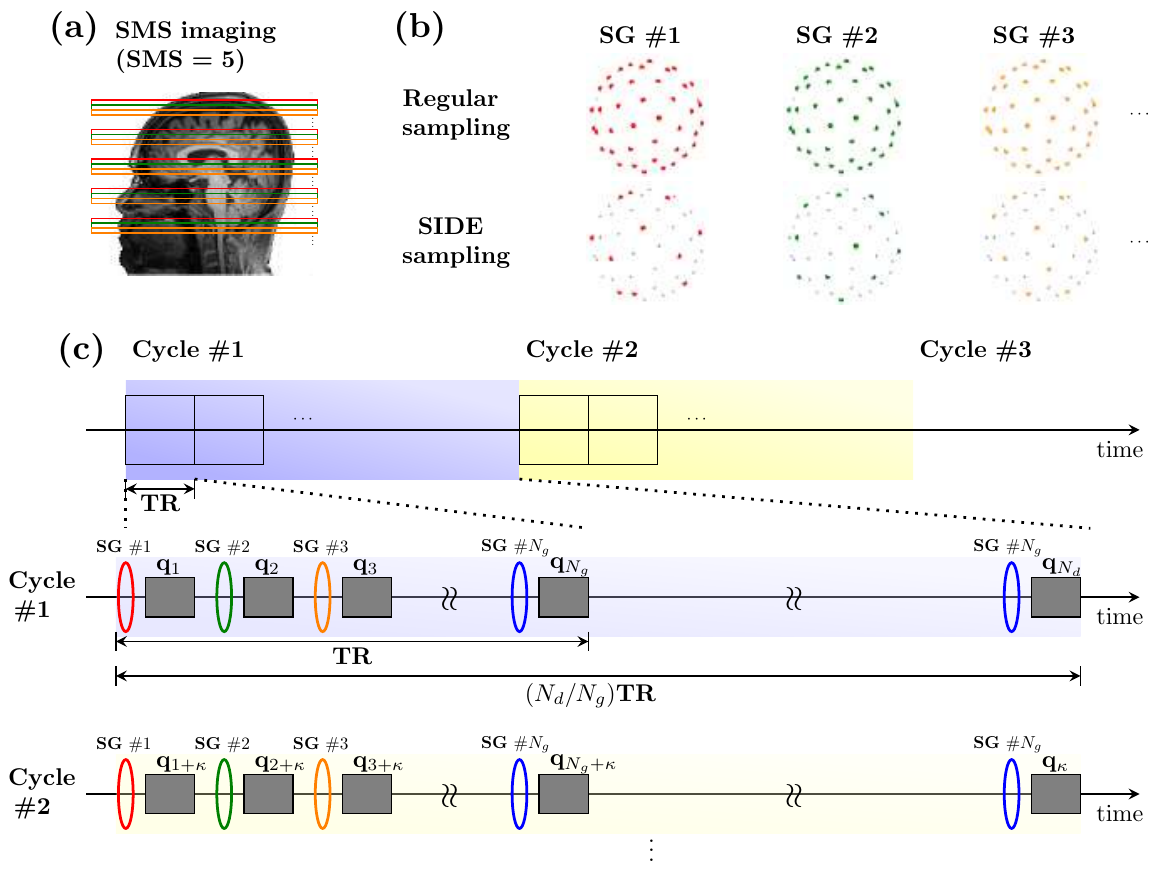}
	\vskip -3ex
	\caption{(a) SMS acquistion with a factor of 5. (b) SIDE acquisition with the SMS slice groups covering different gradient directions, unlike conventional sampling. (c) An illustration of the SIDE sequence.}
	\label{fig:acquisition}
\end{figure*}

Note that in typical pulsed-gradient spin-echo echo planar imaging (PGSE-EPI), the total acquisition time is proportional to the sequence repetition time (TR) and the number of wavevectors. We accelerate the acquisition time via Slice-Interleaved Diffusion Encoding (SIDE) scheme, where only a subset of slices are acquired for each diffusion wavevector. Figure~\ref{fig:acquisition}a shows an example of SIDE acquisition with simultaneous-multislice (SMS) factor 5. Each RF pulse excites a slice group (SG) of 5 slices. In convention diffusion imaging, all SGs in a volume share the same diffusion encoding. In SIDE acquisition, however, each SG may be associated with a different diffusion encoding. 
Figure \ref{fig:acquisition}c illustrates that, in SIDE, the first SMS excitation is followed by the SE-EPI readout block with the first diffusion wavevector $\mathbf{q}_1$ in the gradient table. Likewise, the second SG is encoded by the second diffusion wavevector $\mathbf{q}_2$, and so forth. 
%Such a slice-interleaved diffusion continues across the TRs.
Let $N_g$ denote the number of SGs in a volume and $N_d$ denote the number of wavevectors. To ease implementation, we set $N_d$ as a multiple of $N_g$. The first cycle of diffusion encoding will complete after $N_d/N_g$ TRs (middle row of Figure \ref{fig:acquisition}c).
In the next cycle, the gradient table is offset by $\kappa$ so that the first SG in this cycle is encoded by the $(1+\kappa)$-th gradient direction $\mathbf{q}_{1+\kappa}$ (bottom row in Figure \ref{fig:acquisition}c). $N_g$ cycles cover all the slices of all diffusion wavevectors. A subset of $\tau$ cycles can be selectively acquired for an acceleration factor of $R=N_{g}/\tau$.
%In this work, we do not shorten TR. Instead, we spend the same time for TR by exciting slices applied from different gradients.
%Specifically, the whole slices are divided into $R$ subsets with interleaved slices, where each subset of slices are acquired from different gradients. In this way, the acquisition of dMRI can be accelerated by a factor of $R$ while the spin-history artifacts can be avoided.

The wavevector $\mathbf{q}$ depends on the length, strength, and orientation of the gradient pulses during the measurement sequence and the diffusion time on the pulse length and separation. For PGSE measurements, for example, $\mathbf{q} = \gamma \delta \mathbf{G}$ and $t= \Delta - (\delta/3)$, where $\gamma$ is the gyromagnetic ratio, $\mathbf{G}$ is the diffusion gradient, $t$ is the diffusion time, $\Delta$ is the time between the onsets of the two pulses, and both pulses have length $\delta$. Often we separate $\mathbf{q}$ into a scalar wavenumber $| \mathbf{q} |$ and a diffusion encoding direction $\hat{\mathbf{q}} = \mathbf{q} / |\mathbf{q}|$, which is the direction of the magnetic field gradient in the diffusion-weighted pulses. The $b$-value summarizes both diffusion time and wavenumber $b = t| \mathbf{q} |^{2}$. For spherical acquisition schemes, both $t$ and $|\mathbf{q}|$ are fixed (so $b$ is fixed) and only the gradient direction varies among measurements. The wavevectors can be computed from the $b$-values and gradient directions listed in the gradient table if the diffusion time is known.

\subsection{Reconstruction Problem}

\begin{figure*}[t]
\centering
%\inputTikZ{0.8}{Figures/}{diagram}
\includegraphics[width=0.7\textwidth]{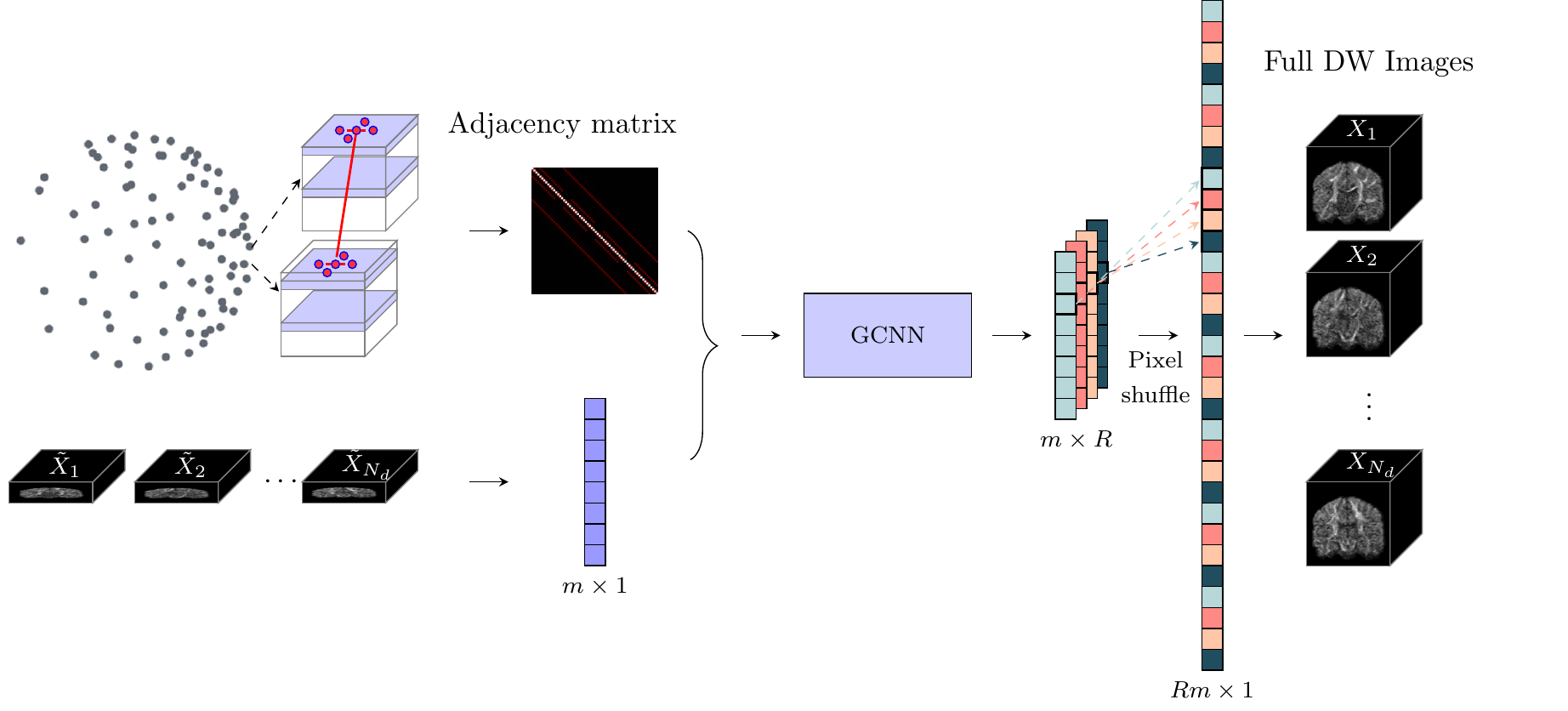}
\vskip -3ex
\caption{GCNN-based reconstruction framework.}
\label{fig:diagram}
\end{figure*}

Our reconstruction method is summarized in Figure~\ref{fig:diagram}.
%Each of the $N$ DW volumes $\{X_n,\, n\,=\,1,\cdots,N\}$ is undersampled in the slice-select direction by a factor of $R$:
For each $n\,=\,1,\cdots,N_d$, let $\tilde{X}_n$ be the slice stacks reorganized from the SIDE volumes acquired with a factor of $R$ in the slice-select direction.
%,  from full DW volume $X_n$. %:
%\begin{equation}
%\tilde{X}_n(\cdot,\cdot,z):= X_n(\cdot,\cdot,Rz+s_n),
%\end{equation}
%where $s_n\in\{0,1,\cdots,R-1\}$ is the slice offset for $X_n$.
Our objective is to reconstruct the full DW volumes $\{X_n\}$ from the undersampled DW volumes $\{\tilde{X}_n\}$ by learning a non-linear mapping function $f$ such that
\begin{equation}\label{eq:obj}
(X_1, \cdots, X_{N_d}) = f(\tilde{X}_1, \cdots, \tilde{X}_{N_d}).
\end{equation}
Instead of reconstructing each DW volume individually, all DW volumes will be simultaneously reconstructed by jointly considering $x$-space and $q$-space neighborhoods. We learn the non-linear mapping function $f$ in \eqref{eq:obj} using a GCNN. By joint consideration of $x$-space and $q$-space  neighborhoods, the complementary information acquired from different DW volumes can be harnessed jointly for effective reconstruction.
Note that the input of the GCNN is a graph constructed from the subsampled DW images instead of the fully interpolated images. This reduces computational complexity and memory requirements.

%\todo{Figure needs to be improved based on our discussion on your RSNA presentation.}

% and the upsampling in slice-select direction is realized by learned interpolation.

\subsection{Graph Representation}
We represent the dMRI signal as a function defined on the nodes of a graph, where each node is determined by a physical spatial location in $x$-space and a wavevector in $q$-space. This graph is encoded with a weighted adjacency matrix $W$, which characterizes the relationships between two nodes. The graph Laplacian operator, defined as $L=D-W$ with $D$ being a diagonal degree matrix, plays an important role in graph signal processing. $L$ can be normalized as $L= I-D^{-1/2}WD^{-1/2}$, where $I$ is an identity matrix. As $L$ is real symmetric positive semidefinite, it has a set of  orthonormal eigenvectors. The eigenvectors of $L$ define the graph Fourier transform that enables the formulation of filtering in the spectral domain \cite{defferrard2016convolutional}. Construction of the adjacency matrix $W$ will be explained in Section~\ref{sec:adjacency}.

\subsection{Spectral Graph Convolution}
Localized graph filters are defined based on spectral graph theory \cite{defferrard2016convolutional}. According to the Parseval's theorem \cite{bronstein2017geometric}, the spatial localization of the convolution corresponds to the smoothness in the spectral domain \cite{bruna2013spectral}. %\todo{Check for correctness.}
Hence, localized filters can be approximated and parameterized by polynomials \cite{bronstein2017geometric}.
Spectral filters represented by the $K$-th order polynomials of the Laplacian are $K$-hop localized in the graph \cite{hammond2011wavelets}.
In the current work, we employ Chebyshev polynomial approximation and define the graph convolutional operation from input $x$ to output $y$ as
\begin{equation}
y=g_\theta (x) = \sum_{k=0}^K \theta_k T_k(\tilde L) x,
\end{equation}
where $T_k(\tilde L)$ is the Chebyshev polynomial of order $k$ evaluated for the scaled Laplacian $\tilde L := 2L/\lambda_{\max}-I$ with $\lambda_{\max}$ being the maximal eigenvalue of $L$. Chebyshev polynomials $\{T_k(\cdot)\}$ form an orthogonal basis on $[-1,1]$ and can be computed by the stable recurrence relation
\begin{equation}
T_k(\lambda) = 2\lambda T_{k-1}(\lambda)-T_{k-2}(\lambda),\text{with}~T_0(\lambda) = 1,~T_1(\lambda) = \lambda.
\end{equation}
%Moreover, by \cite{hammond2011wavelets}, the spectral filters represented by $K$-th order polynomials are exactly $K$-localized.
%The graph convolutional layers in the GCNN can be represented as
%\begin{equation}
%\mathbf{\Phi}^{(l)} = \xi\bigg(\sum_{k=0}^{K}{{\Theta}}_k^{(l)}T_k({\tilde L})\mathbf{\Phi}^{(l-1)}\bigg),
%\end{equation}
%where $\mathbf{\Phi}^{(l)}$ denotes the feature map at the $l$-th layer, ${\Theta}_k^{(l)}$ is a matrix of Chebyshev polynomial coefficients to be learned at the $l$-th layer, and $\xi$ is a non-linear activation function.
Then, the graph convolutional layers in the GCNN can be represented as
\begin{equation}
\mathbf{\Phi}^{(l)} = \xi\bigg(\sum_{k=0}^{K}{{\Theta}}_k^{(l)}T_k({\tilde L})\mathbf{\Phi}^{(l-1)}\bigg),
\end{equation}
where $\mathbf{\Phi}^{(l)}$ denotes the feature map at the $l$-th layer, ${\Theta}_k^{(l)}$ is a matrix of Chebyshev polynomial coefficients to be learned at the $l$-th layer, and $\xi$ denotes a non-linear activation function.

\subsection{Adjacency Matrix}\label{sec:adjacency}
We define the adjacency matrix by jointly considering spatio-angular neighborhoods. The dMRI signal sampling domain can be represented as a graph with each node representing a spatial location $\mathbf{x}_i\in\mathbb{R}^3$ and a normalized wavevector $\hat{\mathbf{q}}_j\in\mathbb S^{2}$. % with corresponding direction $\mathbf{q}_{j} = \hat{\mathbf{q}}_j = \mathbf{q}_j/\|\mathbf{q}_j\|$.
Inspired by the $x$-$q$ space neighborhood matching strategy for dMRI denoising \cite{chen2017neighborhood}, we define a symmetric adjacency matrix $W$ with weight elements $\{w_{i,j;i',j'}\}$:
\begin{dmath}\label{eq:adjacency}
w_{i,j;\,i',j'} {:=} \exp\left(-\frac{\|\mathbf{x}_i - \mathbf{x}_{i'}\|_2^2}{\sigma_{x}^{2}}\right)
\exp\left(-\frac{1-\langle \hat{\mathbf{q}}_j, \hat{\mathbf{q}}_{j'}\rangle^2}{\sigma_{q}^2}\right),
\end{dmath}
where $\sigma_x$ and $\sigma_q$ are the parameters used to control the contributions from the spatial and angular neighborhoods distances, respectively.
We note that the numerators of the arguments of the exponential functions in \eqref{eq:adjacency} are normalized to $[0, 1]$.
%\todo{Did you actually convert the b-vectors to q-vectors?->The vectors in the gradient table are already normalized. So I used them as it is.}
%\todo{See section on the relationship between the $q$-vector and $b$-vector. The direction is the same between the two. The magnitude is related by a square relationship. So it is unclear here whether you are using the $q$-vector or $b$-vector in \eqref{eq:adjacency}.-> I changed to use hat notation.}

\subsection{Graph Convolutional Neural Networks}
 Our generator architecture is based on U-Net \cite{ronneberger2015u} with symmetric encoding and decoding paths. In U-Net, encoding and decoding pathways require pooling and unpooling operations, respectively. Graph coarsening and uncoarsening, which correspond to pooling and unpooling in standard CNNs, are not defined as straightforwardly. For graph coarsening, we adopt the Graclus multi-scale clustering algorithm \cite{dhillon2007weighted} as in \cite{defferrard2016convolutional}. Moreover, a residual convolutional block is employed to ease network training since it can mitigate the problem of vanishing gradients \cite{he2016deep}.
%This fast coarsening algorithm provides coarse graphs at each coarsening level after rearranging the vertices using a binary tree structure \cite{defferrard2016convolutional}.
The graph signal is represented as a single-array vector with permutation indices for rearrangement. The uncoarsening operation is achieved via one-dimensional upsampling operation with a transposed convolution filter \cite{long2015fully}.

Multi-scale input graphs, generated from graph coarsening, are added as new features with graph convolutions at the encoding path of each level.
For skip connection for each level of the encoding path to the decoding path, we apply a transformation module to boost the low-level features to complement the high-level features, as proposed in \cite{nie20183}. Graph convolutions followed by concatenation in the transformation module narrow the gap between low- and high-level features.

%The upsampling operation in slice direction is learned by sub-pixel convolution \cite{shi2016real}, which performs as the standard convolution in low-resolution space followed by a pixel-shuffling operation. The pixel-shuffling operation remaps the input feature maps of size $m\times R$ to the output of size $Rm\times 1$ where $m$ is the number of input graph node.
%The proposed architecture is illustrated in Figure~\ref{fig:architecture}.
The upsampling operation in slice-select direction is realized by standard convolutional layers in the low-resolution space followed by pixel shuffling in the last layer of the network \cite{shi2016real}.
The pixel-shuffling operation converts $R$ pre-shuffled feature maps of size $m\times 1$ to an output feature map of size $Rm\times 1$, where $m$ is the number of input graph nodes and $R$ is the upsampling factor.

In addition to the reconstruction of full DW images, inspired by \cite{chen2019prediction}, we add a branch in the decoding path to compute diffusion indices such as generalized fractional anisotropy (GFA) \cite{tuch2004q}.
% \todo{Cite our RSNA abstrat or arXiv paper when either of them is accepted.}
This branch can help the generator to produce dMRI data with more accurate diffusion indices. In this work, we focus on only GFA, which is estimated by one graph convolutional layer followed by two consecutive fully-connected layers.

\begin{figure*}
\centering

\includegraphics[width=0.8\textwidth]{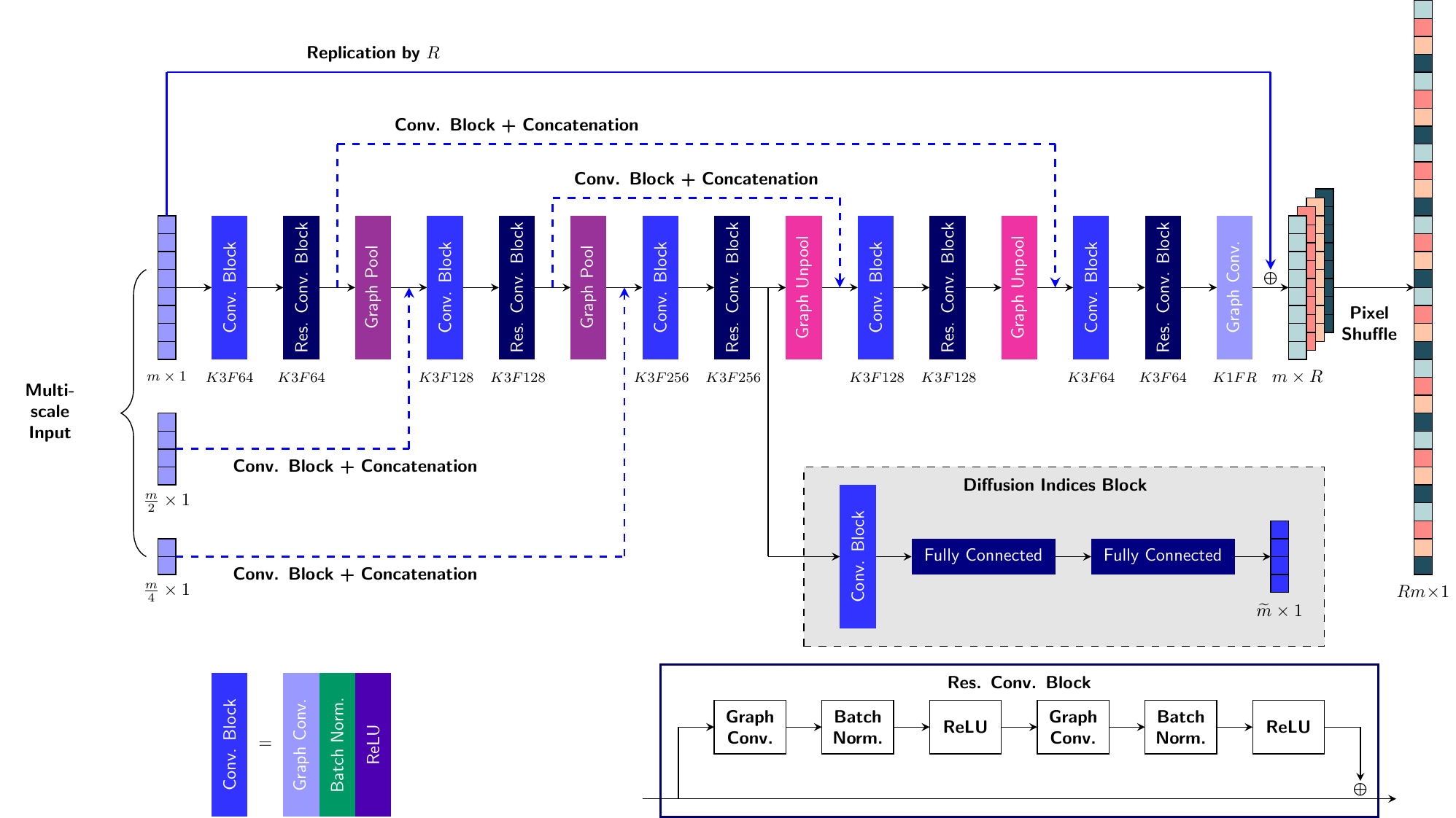}
\vskip -2ex
\caption{The proposed graph CNN architecture. For each convolution layer, $K$ corresponds to Chebyshev polynomial order and $F$ corresponds to the number of feature maps.}
\label{fig:architecture}
\end{figure*}

The architecture of our generator is illustrated in Figure~\ref{fig:architecture}. %For the refinement network, we apply graph convolutional layer followed by two consecutive residual convolutional blocks. %as shown in Figure~\ref{fig:architecture_ref}.
%The final output is obtained via a graph convolution layer with one output channel.
The numbers of feature maps are set to 64, 128, and 256 for the respective levels. The last graph convolutional layer in the image reconstruction branch has $R$ channels with pre-shuffled feature maps for pixel shuffling.

\subsection{Adversarial Learning}
We employ adversarial learning to generate more realistic image outputs. In adversarial learning, the generator attempts to produce outputs that cannot be distinguished from the target images by an adversarially trained discriminator \cite{isola2017image}. The training of the generator and the discriminator is performed in an alternating fashion. Here, the generator is the proposed GCNN with two output branches as shown in Figure \ref{fig:architecture}.
%, and the discriminator $D$ is constructed via patch-GAN \cite{isola2017image} as it is robust and computationally efficient with fewer parameters by removing fully connected layers. The discriminator in patch-GAN is a fully convolutional network, where the discriminator classifies whether the local patch is real or fake, instead of the whole image. 
{We apply two separate discriminators for the predicted DW image and the predicted diffusion index. That is, a discriminator $D_I$ that classifies between the predicted DW image and the real DW image, and another discriminator $D_{\text{GFA}}$ that classifies between the predicted GFA image and the real GFA image.} 
We use leaky ReLU (LReLU) activation for both discriminators with negative slope 0.2, as suggested in \cite{radford2015unsupervised} for stable GANs.

%For adversarial learning, for the input source $\mathbf{x}$ and the target $\mathbf{y}$, we define the discriminator loss as
%\begin{equation}
%\mathcal L_D(\mathbf x, \mathbf y) = \mathcal L_{\text{BCE}}(D(\mathbf y), \mathbf 1) + \mathcal L_{\text{BCE}}(D(G(\mathbf x)), \mathbf 0),
%\end{equation}
%where $\mathcal L_{\text{BCE}}$ is the binary cross-entropy function defined as
%\begin{equation}\label{eq:BCE}
%\mathcal L_{\text{BCE}}({\mathbf{p}}, \mathbf{q}):=-\sum_{i}[q_i\log{{p}_i}+(1-q_i)\log{(1-{p}_i)}].
%\end{equation}
%In (\ref{eq:BCE}), $\mathbf{q}$ takes the values of 1 for real target image and 0 for the generated one, and ${\mathbf p}$ is the predicted probability given by the discriminator.
%
%On the other hand, the generator loss is defined as the combination of pixel-wise difference and adversarial loss:
%\begin{equation}
%\mathcal L_{G_{ADV}}(\mathbf x,\mathbf y)=\lambda_g\|G(\mathbf x) - \mathbf y\|_1 + \lambda_{ADV}\mathcal {L}_{\text{BCE}}(D(G(\mathbf x)), \mathbf 1),
%\end{equation}
%so that the generator $G$ can produce more realistic output to fool the discriminator $D$.
%
%The discriminator consists of three graph convolutions with 64, 128, 256 features, each followed by LReLU and graph pooling.
For the input source $\mathbf{x}$, the target DW image $\mathbf{y}_I$, and the target GFA $\mathbf y_{\text{GFA}}$ , the generator loss is defined as the combination of pixel-wise difference, GFA difference, and adversarial loss:
\setlength{\arraycolsep}{1pt}
\begin{eqnarray}\label{eq:generator}
&~&\mathcal L_{G}(\mathbf x,\mathbf y_I, \mathbf y_{\text{GFA}})\nonumber\\
&=&\lambda_I\|G_I(\mathbf x) - \mathbf y_I\|_1 + \lambda_{\text{GFA}}\|G_{\text{GFA}}(\mathbf x) - \mathbf y_{\text{GFA}}\|_1  \\
&+& \lambda_{\text{ADV}}(\mathcal {L}_{\text{BCE}}(D_I(G_I(\mathbf x)), \mathbf 1) + \mathcal {L}_{\text{BCE}}(D_{\text{GFA}}(G_{\text{GFA}}(\mathbf x)), \mathbf 1)),\nonumber
\end{eqnarray}
%so that the generator $G$ can produce more realistic output to fool the discriminator $D$.
where $\mathcal L_{\text{BCE}}$ is the binary cross-entropy function, and $D_I$ and $D_{\text{GFA}}$ are the discriminators for the predicted images and GFA, respectively.
In \eqref{eq:generator}, $G_I(\mathbf x)$ and $G_{\text{GFA}}(\mathbf x)$ are the outputs of the generator in the image reconstruction branch and diffusion index branch, respectively.
We define the discriminator loss as
\setlength{\arraycolsep}{1pt}
\begin{eqnarray*}
\mathcal L_{D_I}(\mathbf x, \mathbf y_I) &=& \mathcal L_{\text{BCE}}(D_I(\mathbf y_I), \mathbf 1) + \mathcal L_{\text{BCE}}(D_I(G_I(\mathbf x)), \mathbf 0),\\
\mathcal L_{D_{\text{GFA}}}(\mathbf x, \mathbf y_{\text{GFA}}) &=& \mathcal L_{\text{BCE}}(D_{\text{GFA}}(\mathbf y_{\text{GFA}}), \mathbf 1) \nonumber\\
&+& \mathcal L_{\text{BCE}}(D_{\text{GFA}}(G_{\text{GFA}}(\mathbf x)), \mathbf 0).
\end{eqnarray*}
%where $D_I$ and $D_{\text{GFA}}$ are the discriminators for the predicted images and GFA, respectively.
$D_I$ consists of three graph convolutions with 64, 128, 256 features, each followed by LReLU and graph pooling. $D_{\text{GFA}}$ consists of three fully-connected layers with 64, 32, and 1 node(s), respectively.

\section{Experimental Results}\label{sec:experiment}
The proposed method is validated with retrospectively undersampled HCP data and in vivo human brain data acquired with SIDE.

\subsection{Materials}
\subsubsection{Simulated Data}
We demonstrate the effectiveness of our method using randomly selected 16 subjects from the Human Connectome Project (HCP) database \cite{sotiropoulos2013advances}. We perform 4-fold cross-validation with 12 subjects for training and 4 subjects for testing at each fold. Each subject has a total of 270 DW images of voxel size $1.25^3 \,\text{mm}^3$, 90 each for $b=1000,\,2000,\, 3000\, \text{s/mm}^2$. We retrospectively undersampled the images by factors $R=3, 4, 5$ and $6$. Specifically, the set of DW images was divided into $R$ subsets so that the wavevectors were uniformly distributed in each subset. For each subset, the source images were generated by undersampling the original images with a slice offset. For each undersampling factor $R$, we extract input and output patches with size $R\times R\times 1 \times 90$ and $R \times R\times R \times 90$, respectively. 
%In GCNN, as the spatial patch size increases, the computational complexity increases due to the increased number of graph nodes. For high undersampling factor $R=6$, we set the input and output patches as $5\times 5\times 1\times 90$ and $5\times 5\times 6\times 90$, respectively. 
In order to harness more contextual information to recover missing slices, the patch size increases with the undersampling factor $R$. 
%For orthogonal SR, we first divide 90 DW images into 3 scan directions so that each scan direction has 30 unique gradient vectors.
%\todo{need to add a figure to show the distribution of wavevectors?}
%\todo{Why does the patch size change with $R$? (A) There's no specific rational for the selection of patch size. I simply set $R\times R\times 1\times 90$, in general. But for $R=6$, since the number of nodes was increased, resulting high computational cost, I used the reduced patch size. }

\subsubsection{SIDE Data}
After obtaining informed consent, we acquired dMRI data from seven healthy subjects using a protocol approved by the institute and a 3T Siemens whole-body Prisma scanner (Siemens Healthcare, Erlangen, Germany). Diffusion imaging was performed with a monopolar diffusion-weighted PGSE-EPI sequence. The SMS RF excitation with controlled aliasing (blipped-CAIPI) was employed to reduce the penalty of geometry factor (g-factor). The SMS factor is 5. Imaging parameters were as follows: resolution= $1.5^3 \,\text{mm}^3$; FOV $= 192\times 192 \times 150\,\text{mm}^3$; image dimensions $= 128\times 128\times 100$; partial Fourier = 6/8; no in-plane acceleration was used; bandwidth = 1776 Hz/Px; 160 wavevectors distributed over the 4 b-shells of $b = 500, 1000, 2000$, and $3000\,\text{s/mm}^2$ with 16, 32, 48, and 64 non-collinear directions respectively, plus one $b=0\,\text{s/mm}^2$ scan; TR/TE$~=~3120/90$\,ms; 32-channel head array coil. The total acquisition time for full DW images is 8 mins and 19 secs for each phase-encoding direction. Note that $N_{d}=160$ and $N_{g}=100/5=20$. SG is shifted by $\kappa=1$ for each cycle.

We performed leave-one-out cross-validation for 48 DW images ($b=2000\, \text{s/mm}^2$) with undersampling factors $R=2, 4, $ and $10$. For each undersampling factor $R$, we extract input and output patches with size $5\times 5\times 1 \times 48$ and $5 \times 5\times R \times 48$, respectively. For $R=2$, we selected the 1st to 10th cycles from 20 cycles. For $R=4$, we selected the 1st, 3rd, 5th, 7th, and 9th cycles. For $R=10$, we selected the 1st and 16th cycles. The ground truth full DW images were  acquired with all 20 cycles.

\subsection{Implementation Details}
All DW images were normalized by their respective non-DW image ($b_0$). The weight controlling parameters in (\ref{eq:adjacency}) are set to $\sigma_x^2 = 0.1$ and $\sigma_q^2 = 1.0$  for joint consideration of spatial and angular distances. The order of the Chebyshev polynomials $K$ is set to $3$. For the loss functions, we set $\lambda_I = 1.0$, $\lambda_{\text{GFA}} = 0.1$, and $\lambda_{\text{ADV}} = 0.01$. The proposed method was implemented using TensorFlow 1.13.1 and trained with the ADAM optimizer with an initial learning rate of 
%0.0001
1$\times$10\textsuperscript{-4} 
and 
%0.00001 
1$\times$10\textsuperscript{-5} 
for the generator and the discriminators, respectively, and a mini-batch size of 10. The learning rate is decreased with an exponential decay rate of 0.95 at every 10,000 steps.

\subsection{Results}
\subsubsection{Simulated Data}
We compared our method (with and without GFA loss) with bicubic interpolation and 3D U-Net \cite{cciccek20163d} applied to input images upsampled via bicubic interpolation, for four different undersampling factors and three different b-shells. %As the results from bilinear interpolation are slightly worse than bicubic interpolation, we omit the bilinear interpolation results.
For 3D U-Net, we extracted patches of size $16\times 16\times 16$ with patch offset $8$.
We also implemented a GCNN method which is applied to the patches extracted from the upsampled inputs by bicubic interpolation as in 3D U-Net. This Bicubic+GCNN is essentially the same as the proposed method except the last pixel-shuffling layer. The patch size was fixed as $4\times 4\times 4\times 90$ with patch offset $3$ for the Bicubic+GCNN for all $R$'s. For our method, we set the input and output patch sizes as $R\times R\times 1\times 90$ and $R\times R\times R\times 90$, respectively. 
%The width/height of the patch was set to $R$ for $R< 6$ and $5$ for $R=6$. 
Table~\ref{tab:no_param} gives the number of training parameters of the generator for $R=4$. Note that the number of channels in Bicubic+GCNN was set to 64 for all layers so that the number of parameters was comparable to the other methods.

\setlength{\tabcolsep}{5pt}
\begin{table}
\centering
\caption{Number of training parameters ($R=4$).}
\label{tab:no_param}
\vskip -2ex
\begin{tabular}[\textwidth]{@{}lcccc@{}}
\toprule
~ & \multicolumn{1}{c}{Bicubic+3D U-Net} & \multicolumn{1}{c}{Bicubic+GCNN} & \multicolumn{1}{c}{GCNN} & \multicolumn{1}{c}{GCNN+GFA loss}\\
\midrule
~ & $5.60\times 10^6$ & $6.63\times 10^6$ & $5.37\times 10^6$ & $5.43\times 10^6$\\
\bottomrule
\end{tabular}
\end{table}

\begin{figure*}
\centering
\includegraphics[width=0.9\textwidth]{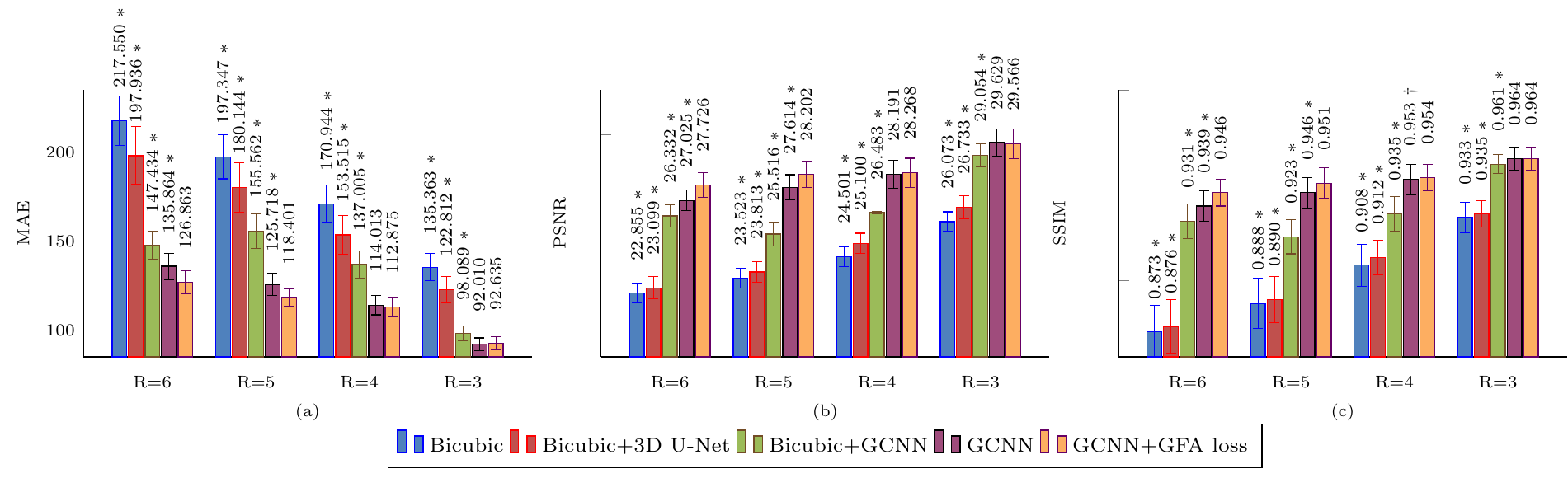}
\vskip -2ex
\caption{Quantitative comparison of DW images using (a) MAE, (b) PSNR, and (c) SSIM, under different undersampling factors. $\ast$ and $\dagger$ indicate the $p$-value $< 0.01$ and $<0.05$, respectively, compared to GCNN+GFA loss. }
\label{fig:quant_dwi}
\end{figure*}

\begin{figure*}
\centering
\includegraphics[width=0.9\textwidth]{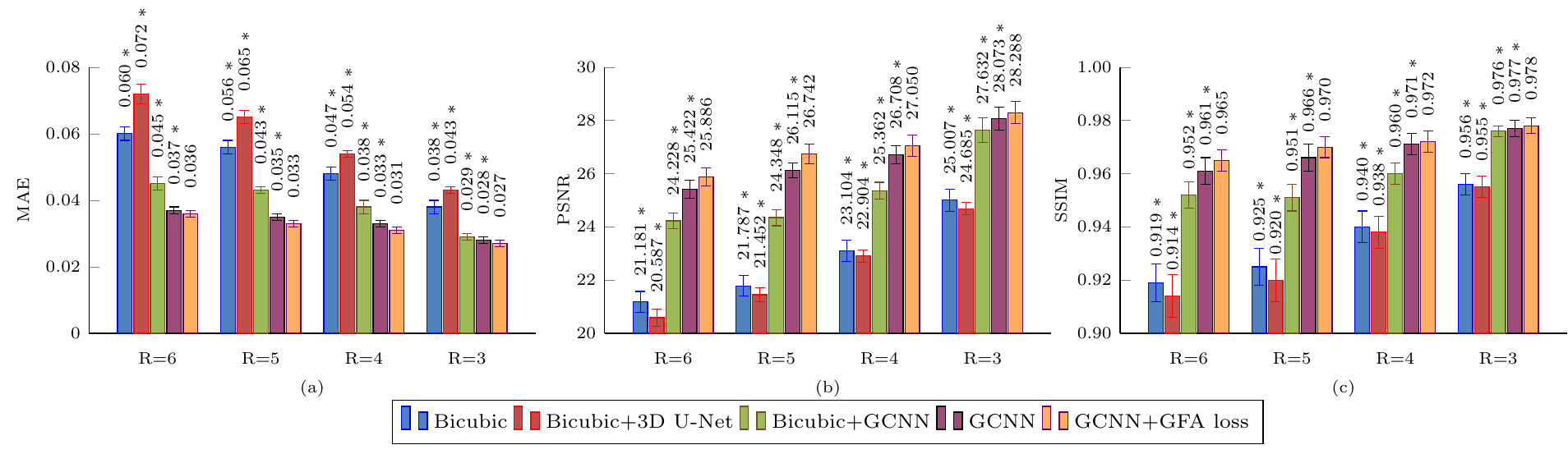}
\vskip -2ex
\caption{Quantitative comparison of GFA using (a) MAE, (b) PSNR, and (c) SSIM, under different undersampling factors. $\ast$ and $\dagger$ indicate the $p$-value $< 0.01$ and $<0.05$, respectively, compared to GCNN+GFA loss. }
\label{fig:quant_GFA}
\end{figure*}

\begin{figure*}
\centering
\includegraphics[width=0.85\textwidth]{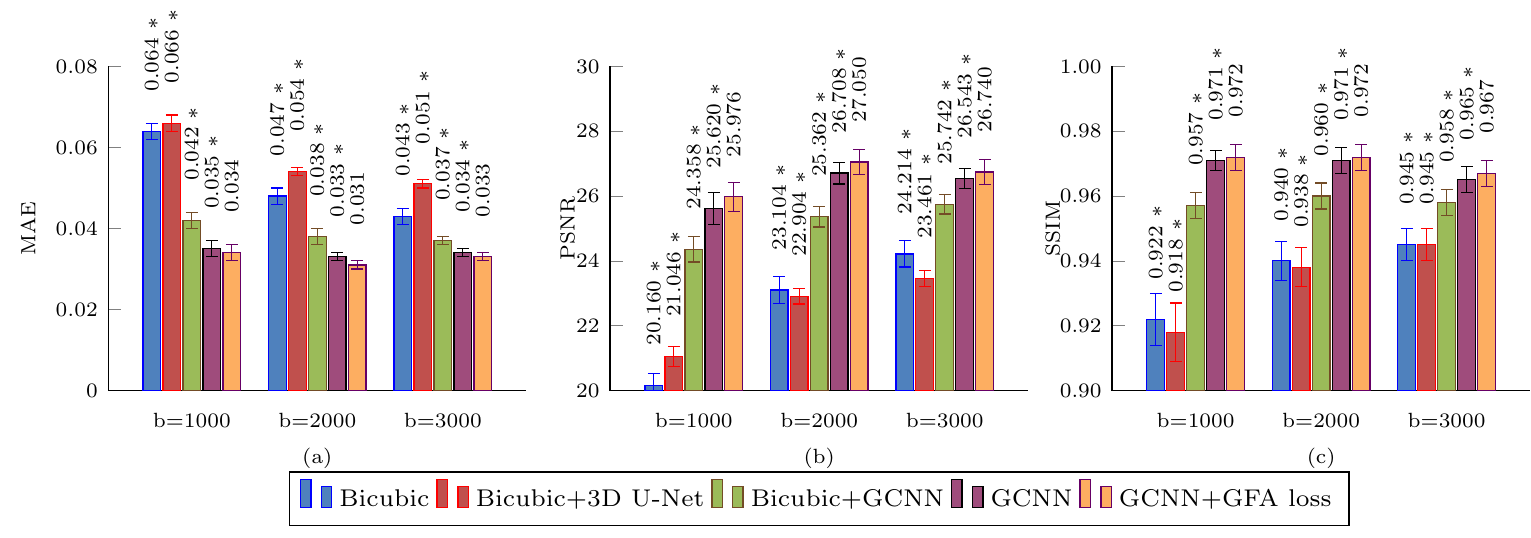}
\vskip -2ex
\caption{Quantitative comparison of GFA using (a) MAE, (b) PSNR, and (c) SSIM, under different shells. $\ast$ and $\dagger$ indicate the $p$-value $< 0.01$ and $<0.05$, respectively, compared to GCNN+GFA loss.}
\label{fig:quant_GFA_bvalues}
\end{figure*}

We measure the reconstruction accuracy of DW images by means of mean absolute error (MAE), peak signal-to-noise ratio (PSNR), and structural similarity index (SSIM). We also computed the GFA maps of the reconstructed DW images. The quantitative results under the different undersampling factors at fixed $b\,=\,2000\, \text{s/mm}^2$ are summarized in Figure~\ref{fig:quant_dwi} for DW images and in Figure \ref{fig:quant_GFA} for GFA maps. We also compared the proposed method for different b-values at $R=4$ and the results are summarized in Figure \ref{fig:quant_GFA_bvalues}. %Note that GCNN and 3D U-Net are applied for the same input patches. 
The quantitative results demonstrate that GCNN is superior to 3D U-Net as it exploits angular neighborhood information in the form of graphs. Our method yields results that are better than Bicubic+GCNN, since Bicubic+GCNN uses fixed hand-crafted upsampling, whereas our method learns the upsampling mapping via the pixel-shuffling operation. Moreover, our method is faster than Bicubic+GCNN with lower computational cost.
%We further conducted the Wilcoxon signed-rank test to validate whether the improvement of the proposed method compared to other methods is statistically significant. 
The results in Figures~\ref{fig:quant_dwi}, \ref{fig:quant_GFA} and \ref{fig:quant_GFA_bvalues} indicate that the improvements over other competing methods are statistically significant ($p<0.01$, Wilcoxon signed-rank test).
%The results in Figure~\ref{fig:quant_GFA} indicate that the improvements over 3D U-Net are statistically significant ($p<0.05$) for all except one case.

%The quantitative results are summarized in Figure \ref{fig:mae}. The normalized root-mean-square errors (RMSE) in terms of spherical harmonic (SH) coefficients up to order 8 are summarized in Table \ref{tab:shc}.

\begin{figure*}
\centering

\includegraphics[width=0.75\textwidth]{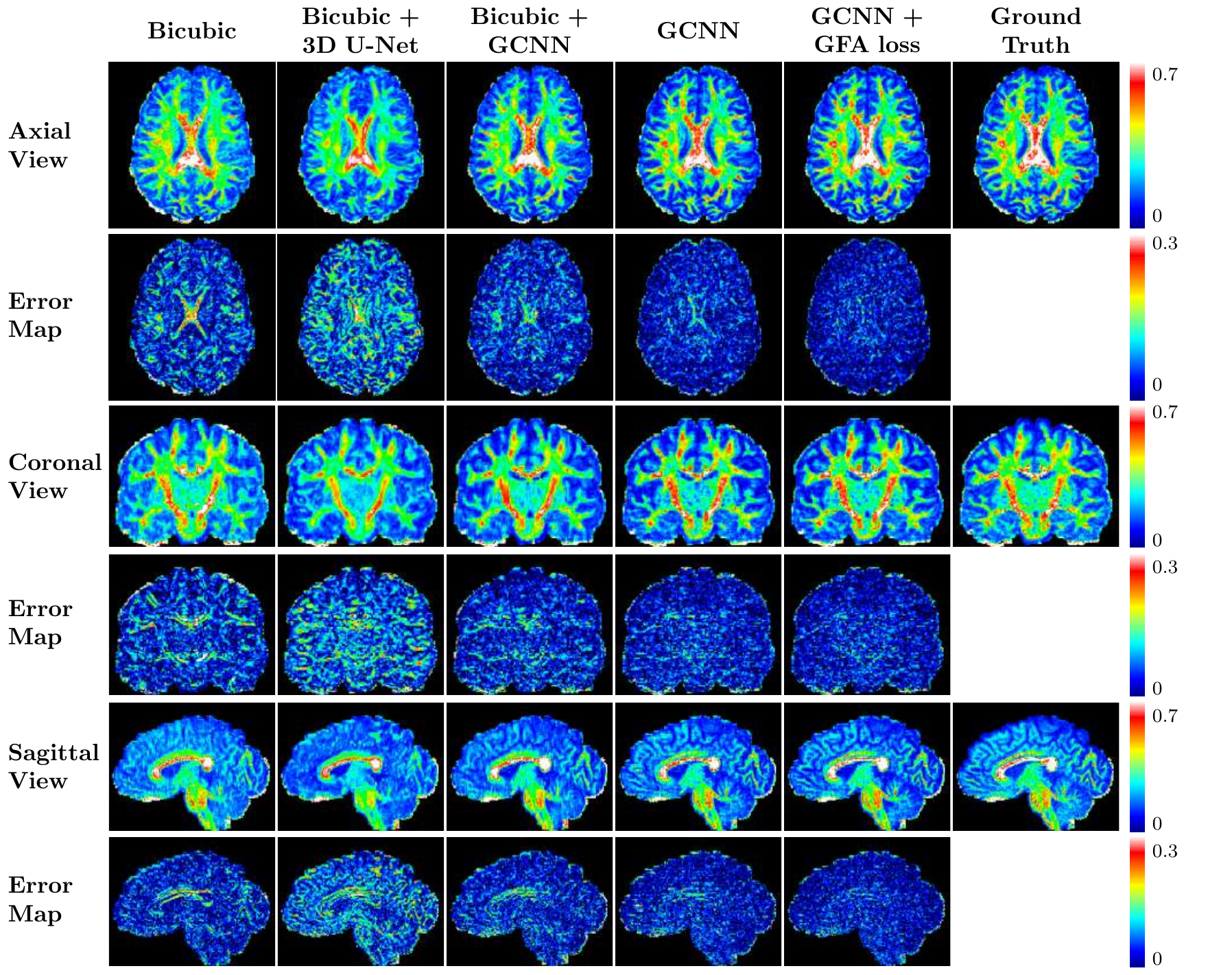}
\vskip -2ex
\caption{Computed GFA maps from the predicted DW images and the corresponding error maps shown in multiple views ($R=5, b=2000$).}
\label{fig:gfa}
\end{figure*}

\begin{figure*}
\centering

\includegraphics[width=0.7\textwidth]{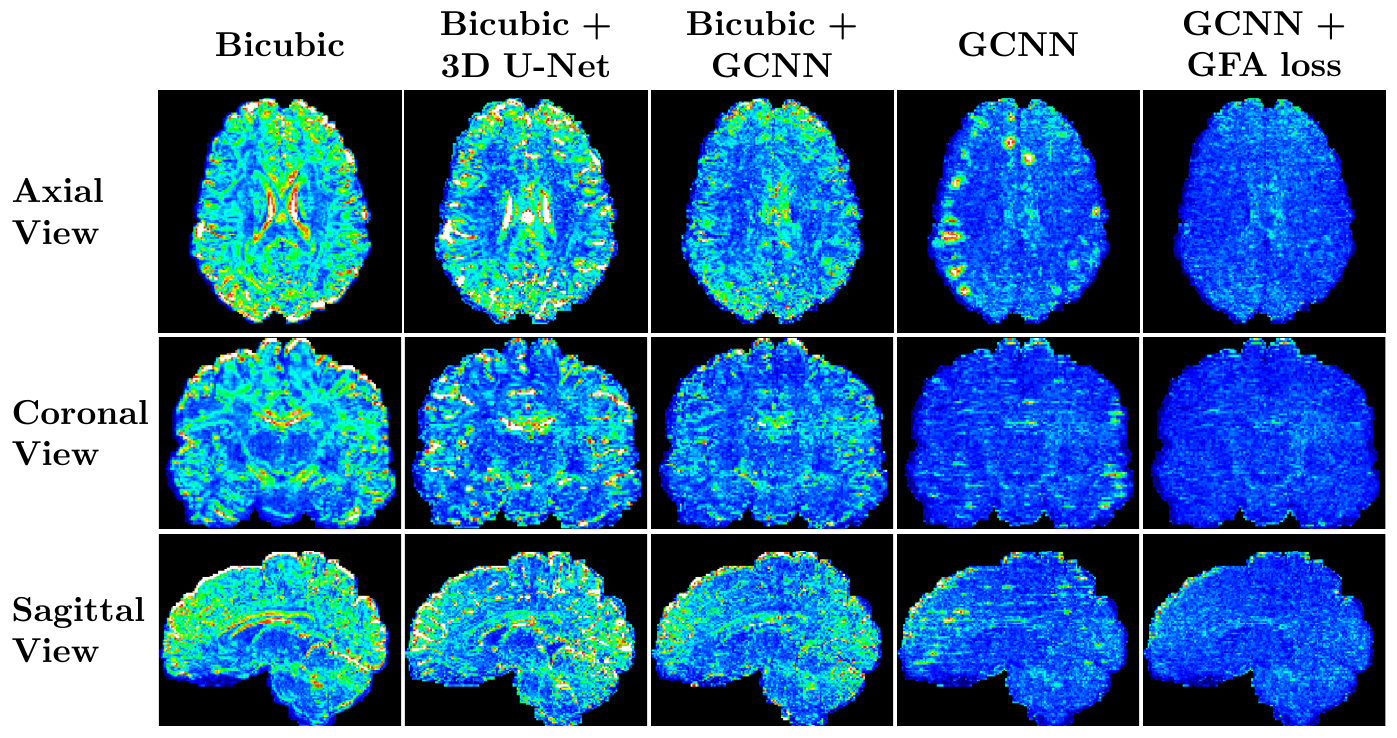}
\vskip -2ex
\caption{RMSE of SH coefficients shown in multiple views with display window $[0, 300]$ ($R=5, b=2000$).}
\label{fig:sh}
\end{figure*}

Representative reconstruction results for GFA at $R=5$ and $b=2000\, \text{s/mm}^2$, shown in Figure~\ref{fig:gfa}, indicate that the proposed methods recover more structural details compared with the competing methods. Moreover, the error in slice-select direction
%\todo{Why are you explicitly mentioning error in `slice-select direction' here? -> I wanted to emphasize the advantage of GFA loss since we observed the error in slice-select direction without GFA loss.}
is significantly reduced when GFA loss is considered. Figure~\ref{fig:sh} shows root-mean-square-error (RMSE) map of spherical harmonic (SH) coefficients of maximum order 8.

\begin{figure*}
\centering
\includegraphics[width=0.90\textwidth]{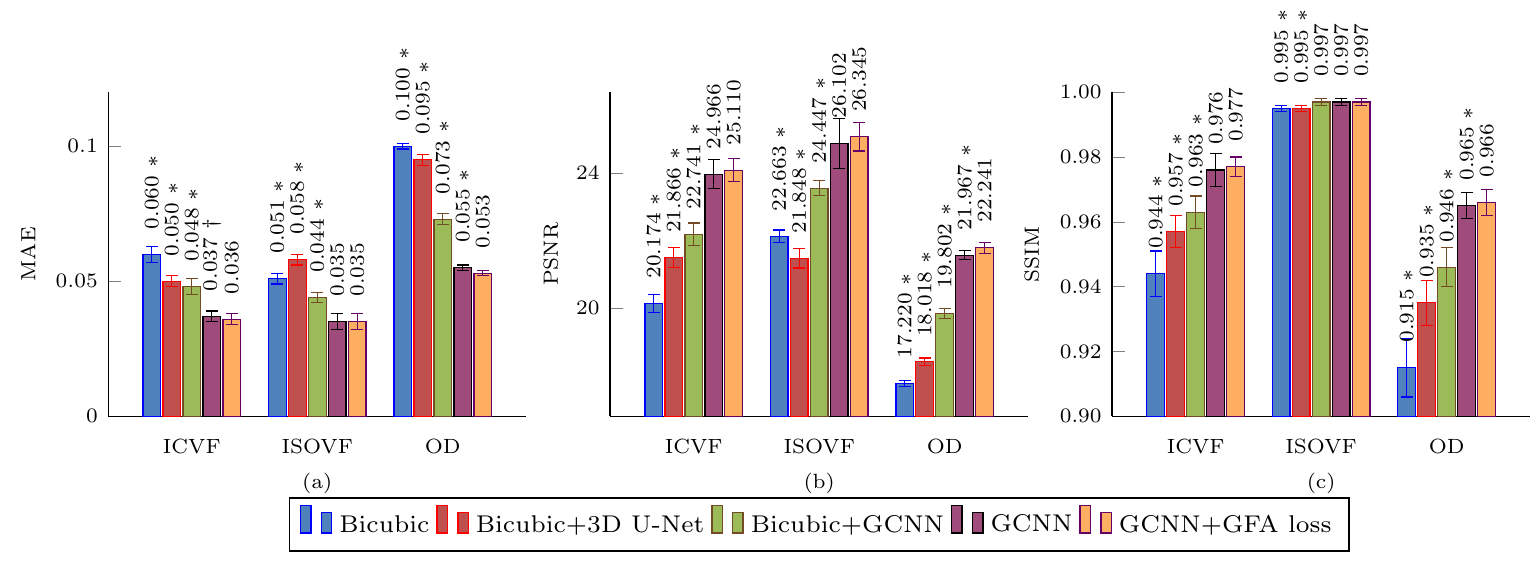}
\vskip -2ex
\caption{Quantitative comparison of the measures of NODDI models using (a) MAE, (b) PSNR, and (c) SSIM. $\ast$ and $\dagger$ indicate $p < 0.01$ and $p <0.05$, respectively, compared to GCNN+GFA loss.}
\label{fig:quant_noddi}
\end{figure*}

\begin{figure*}
\centering

\includegraphics[width=0.75\textwidth]{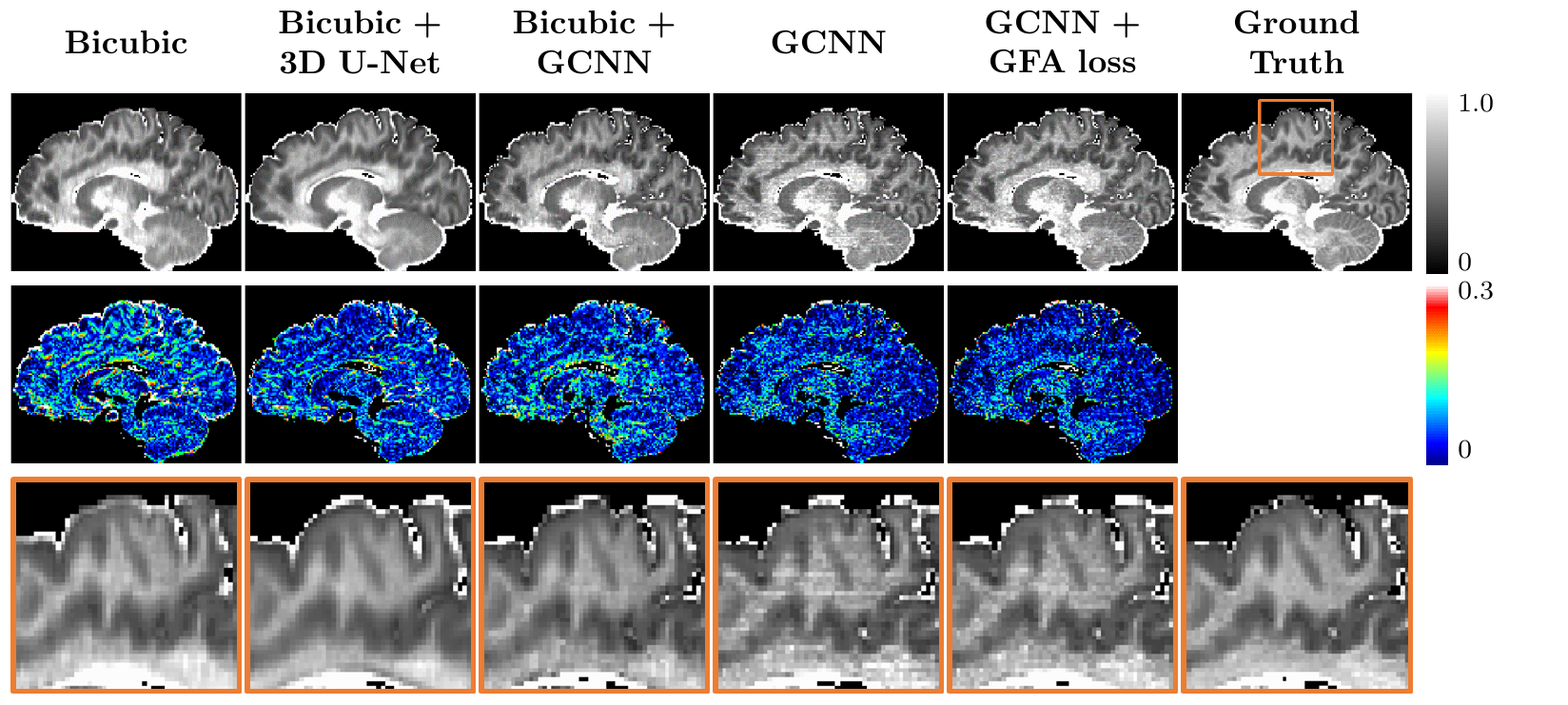}
\vskip -2ex
\caption{Representative ICVF maps (top row), the corresponding error maps (middle row), and their close-up views (bottom row).}
\label{fig:icvf}
\end{figure*}

Evaluation was also performed based on the diffusion indices given by neurite orientation dispersion and density imaging (NODDI) \cite{zhang2012noddi}, applied on multi-shell data with each shell predicted using the different methods. 
%data in each shell and merging the results to become multi-shell data. 
The quantitative results for intra-cellular volume fraction (ICVF),  isotropic volume fraction (ISOVF), and orientation dispersion (OD) are summarized in Figure~\ref{fig:quant_noddi}. 
%As the estimation of ICVF and OD is degenerate for region containing only CSF \cite{ye2019deep}, we excluded CSF regions for ICVF and OD comparison. 
A representative result of ICVF for $R=4$ is shown in Figure~\ref{fig:icvf}. While either the bicubic interpolation or the results learned from fixed interpolated source are blurry, our method yields results that are relatively sharp and closer to the ground truth. Moreover, the error shown in slice-select direction is significantly reduced with GFA loss.

%Figure \ref{fig:fodf} shows that our method can yield fiber orientation distribution functions (ODFs) that are closer to the ground truth with less partial volume effects highlighted in the rectangles.

%
%We also extracted four representative tract bundles from whole brain tractography using the multi-ROI approach described in \cite{mori1999three}. We extracted the forceps major (FMajor) using ROIs drawn in the occipital cortex and corpus callosum (CC), and also the forceps minor (FMinor) using ROIs drawn in the prefrontal cortex and CC. For the corticospinal tract (CST), ROIs are drawn in precentral gyrus and posterior limb of the internal capsule. For cingulum cinulate gyrus part (CGC), ROIs are drawn in the middle of splenium of the CC and the middle of genu of CC. Figure \ref{fig:tractography} shows that our method can yield richer fiber tracts that better resembles the ground truth. Some fiber tracts are missing in the interpolated results.
%\begin{figure*}
%\centering
%\inputTikZ{1}{Figures/}{tract}
%%\vskip -2ex
%\caption{Representative tractography results ($R=4$).}
%\label{fig:tractography}
%\end{figure*}

\begin{figure*}
\centering
\includegraphics[width=0.8\textwidth]{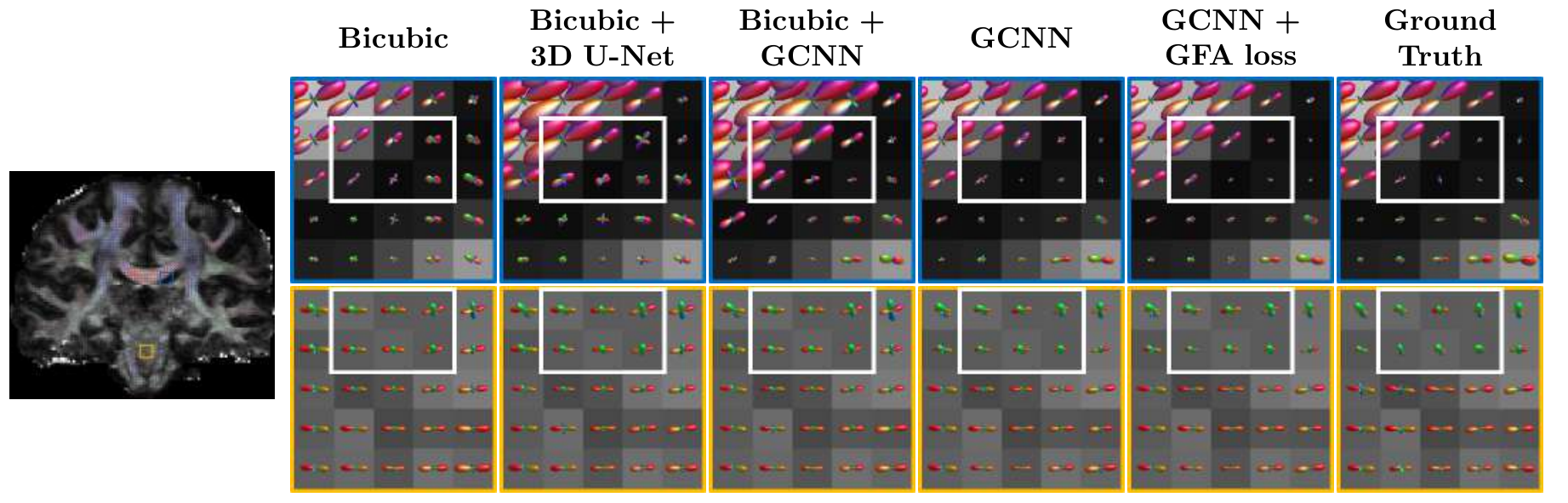}
\vskip -2ex
\caption{Representative fiber ODFs ($R=4$).}
\label{fig:fodf}
\end{figure*}

Figure \ref{fig:fodf} shows that our method provides more coherent and accurate fiber orientation distribution functions (ODFs) with less partial volume effects, especially in the regions marked by the rectangles.

\subsubsection{SIDE Data}
\begin{figure*}
	\centering
	\includegraphics[width=0.8\textwidth]{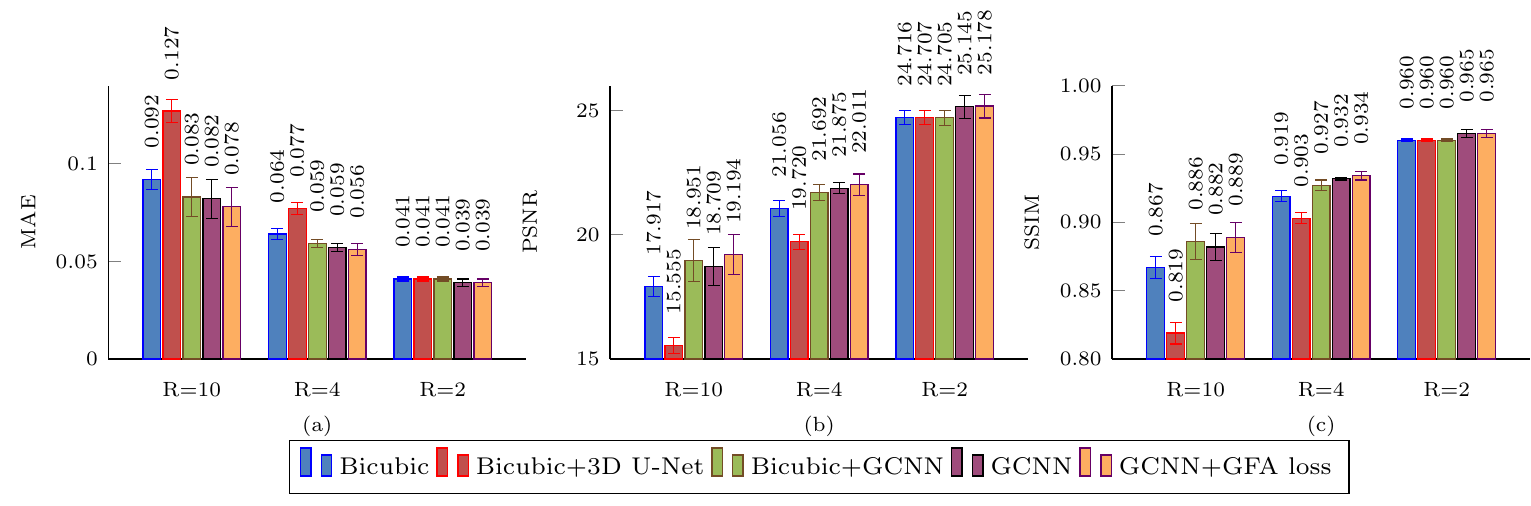}
	\vskip -2ex
	\caption{Quantitative comparison of GFA using (a) MAE, (b) PSNR, and (c) SSIM, under different undersampling factors.}
	\label{fig:quant_side}
\end{figure*}

\begin{figure*}
	\centering
	\includegraphics[width=0.75\textwidth]{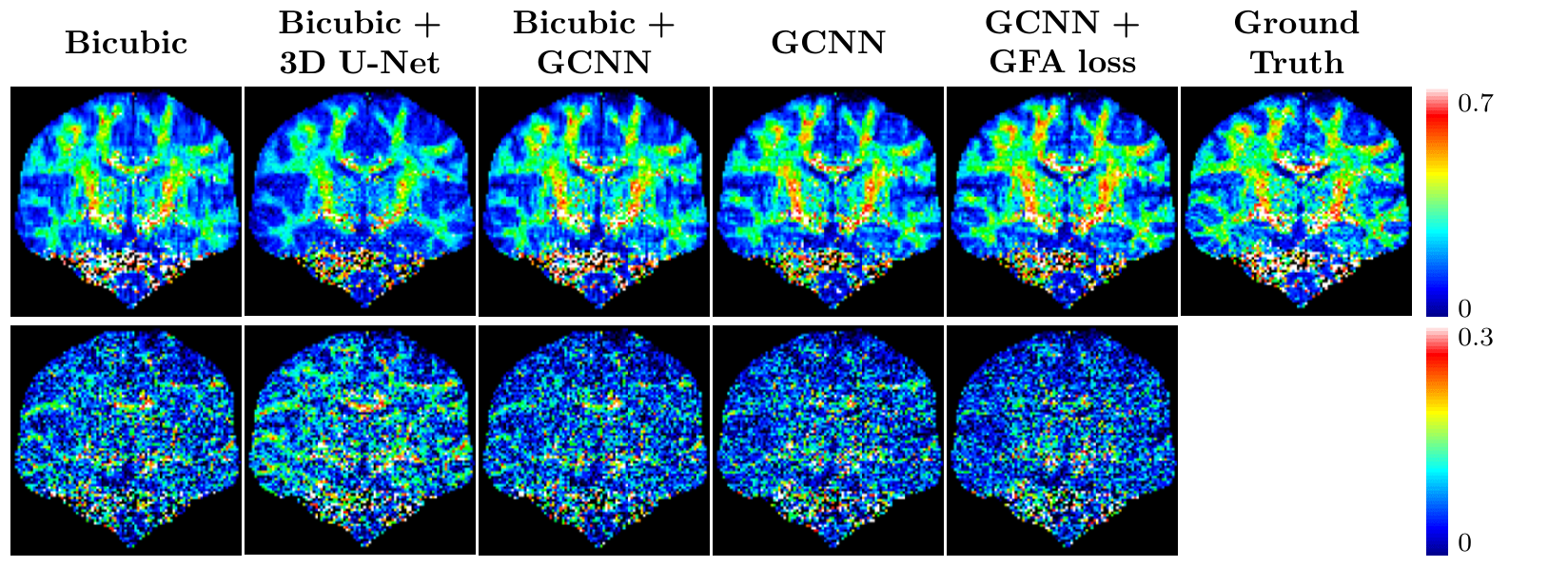}
	\vskip -2ex
	\caption{Computed GFA maps from the predicted DW images and the corresponding error maps shown in coronal views ($R=4$).}
	\label{fig:gfa_side}
\end{figure*}

We also evaluated our method with the SIDE data of seven subjects for $b\,=\,2000\, \text{s/mm}^2$ under various acceleration factors $R\,=\,2,\,4$, and $10$. The patch sizes for 3D U-Net and Bicubic+GCNN were set to be the same as the simulated data. The quantitative results for the different undersampling factors are summarized in Figure~\ref{fig:quant_side} for GFA maps, again demonstrating that our method is superior to the methods exploiting only spatial neighborhood information. Note that for $R\,=\,2$, the difference between our method and other competing methods are marginal. However, the improvement becomes more significant as the undersampling factor increases.
%\todo{how much percentage? p value?-> I didn't calculate p-values for SIDE data since it has only 7 dataset. Do you think it's meaningful to report p-values for such a small dataset?}. 
Representative reconstruction results for GFA at $R=4$, shown in Figure~\ref{fig:gfa_side}, confirm that the proposed methods recover more structural details, especially in the body of corpus callosum and near the cerebral cortex, compared with the competing methods.
%\todo{quantitative comparisons need to be provided -> Figure 12 shows quantitative comparison in terms of GFA maps. Do we need more results?}. 

Note that we selected certain cycles out of 20 cycles for different undersampling factors so that the acquired slices for each wavevector are as equally-spaced as possible. For the HCP simulated data, the undersampled slices were exactly equally-spaced.
 
%This is due to the sub-optimal design of the current SIDE data and will be discussed in \ref{sec:discuss}.
  
%\todo{This section is too short. It is the most important results section. Maybe add more text to more clearly describe the observations and conclusions.}

\section{Discussion}\label{sec:discuss}

We have shown that the proposed GCNN is capable of recovering DW images from highly undersampled data acquired via SIDE with acceleration factor as high as 10 (50 with SMS acquisition). We will discuss here some limitations of the proposed framework and future work for improvements.

%\subsection{Limitations}
Gradient ordering throughout the cycles needs to be further optimized. Currently, gradient ordering is varied at each cycle by offsetting the gradient table with a step of 1. This is not necessarily optimal and causes the slices acquired in two consecutive cycles to be highly correlated. That is, adjacent slices for each gradient are acquired as the cycle progresses. A more optimal ordering strategy should ensure that incoherent information is covered across cycles. Incoherence in general promotes more effective recovery of unsampled information. 

The proposed reconstruction framework assumes a consistent gradient table in training and testing, preventing the learned mapping to be applied to data collected using different gradient tables. This limitation can be overcome by utilizing GCNN methods that are based on heterogeneous graphs \cite{such2017robust}. Another potential solution is to learn the mappings between representations, such as spherical harmonics (SHs), instead of the signal directly. That is, the mapping between the SHs of the undersampled data to the SHs of fully-sampled data can be learned and applied to data collected with different numbers of gradient directions as long as the signal is consistently represented by SHs up to a fixed maximum order.

Conventional dMRI acquisition typically covers the slices associated with each diffusion gradient fully before proceeding with the next gradient. 
In contrast, SIDE acquisition covers the slice groups of all gradients in each cycle and repeats in the next cycle with a gradient offset (see Figure~\ref{fig:acquisition}c). 
In other words, each cycle covers partial information of all gradients. 
This is the key characteristic of SIDE that allows unsampled information to be recovered from the data acquired in a few cycles.

The above observation has important implications on scan disruption caused for example by subject motion.
In conventional dMRI acquisition, disruption results in the total loss of information with respect to some diffusion gradients, which can be challenging to recover. 
On the other hand, disruption of SIDE acquisition results in partial loss of slices of a volume. The lost slices can be recovered from adjacent slices acquired with the same and similar gradients.

%Motion correction can implemented for SIDE data in a manner that is different from conventional acquisitions. 
SIDE acquisition can potentially improve post-acquisition motion correction, for example, by registering motion-affected data to motion-free data. If SIDE acquisition is able to complete motion-free for a few cycles, the acquired data, which cover all gradients, can be used as a reference for registration-based motion correction.
Alternatively, a reference for registration-based correction can be constructed by generating the full data using the motion-free data using our reconstruction framework.

%For example, if abrupt motion happens in the middle of the scan, the images acquired 
%Unlike conventional dMRI data, the pre-motion SIDE data contain information associated with all gradients, providing a good reference onto which post-motion data can be registered. 

The time saved by using SIDE acquisition can be used for a denser coverage of the $q$-space, which in turn also improves reconstruction based on the SIDE data since more information in the $q$-space can be leveraged for data prediction. This opens the opportunity for dense $q$-space coverage to allow for better prediction of tissue microstructure and white matter pathways. The balance between $q$-space coverage and $x$-space coverage in terms of slices is a subject of future research.

% Discuss different slice-select directions?

%\rev{
%Typical diffusion imaging required the entire scan to acquire a complete set of diffusion directions. For SIDE acquisition, however, the acquisition of all the diffusion directions can be finished in one cycle, as shown in Figure \ref{fig:acquisition}c. 
%In the case that the diffusion scan is terminated unexpectedly, the SIDE approach might have acquired the entire set of diffusion directions multiple times, offering a good chance to reconstruct the unacquired images. This feature of SIDE is particularly useful for pediatric imaging because the baby may not stay incorporated early before the end of the scan.
%}

%(Any limitation? Potential concern?)

%(detailed discussion of the results?)
%{To increase the reconstruction performance, we added a branch to predict diffusion indices. Our hypothesis is that this additional prediction can leverage the reconstruction of DW images for more accurate diffusion index parameters. We showed the feasibility of predicting only GFA; however, other kinds of diffusion index can be used and multiple parameters can be simultaneously predicted to help reconstruction of DW images.}

%Future work will investigate the use of high-resolution structural image priors\todo{what makes you believe the structural information will help? Has anyone incorporated the anatomical information into the recon of dMRI? If so, citation?}. \todo{please elaborate the following sentence}\rev{For example, T1- or T2-weighted image guided training may provide more information and enable improved SR.}

\section{Conclusion}\label{sec:conclusion}
We have proposed a novel slice-based sampling technique to accelerate dMRI acquisition. The slices in a volume are encoded with multiple diffusion gradients to allow for rapid sampling of information associated with different gradients.
Highly subsampled data acquired using this approach can then be fed into a deep learning framework that jointly considers spatial and wavevector domains to reconstruct the full data. The high acceleration factor achievable opens the door to future opportunities for improved motion correction and dense $q$-space imaging.

%\todo{the following sentence is hard to understand. Please rephrase}\rev{Each DW image is subsampled with a different slice offset so that the complementary information captured by each DW image can be used to improve the reconstruction of the other DW image.}
%Each DW image is undersampled with a different slice offset and the missing slices are reconstructed by exploiting neighborhood information in the spatial and angular domains. 
%We formulate the reconstruction as an image synthesis problem. The non-linear mapping from the subsampled to fully-sampled DW images was learned using GCNN. \todo{not sure what it means}\rev{Spatio-angular relationship is jointly considered when encoding dMRI signal as graphs for the GCNN}. We demonstrate that the proposed method outperforms \todo{"the competing methods" is not the right description. Please simply describe the methods that were in comparison}\rev{the competing methods}. 
%In addition, we have shown that the method with GFA loss recovers more structural details.
%The experimental results demonstrate that the proposed method outperforms two commonly used interpolation methods.

\bibliographystyle{IEEEtran}
\bibliography{reference}

% Generated by IEEEtran.bst, version: 1.12 (2007/01/11)
\begin{thebibliography}{10}
\providecommand{\url}[1]{#1}
\csname url@samestyle\endcsname
\providecommand{\newblock}{\relax}
\providecommand{\bibinfo}[2]{#2}
\providecommand{\BIBentrySTDinterwordspacing}{\spaceskip=0pt\relax}
\providecommand{\BIBentryALTinterwordstretchfactor}{4}
\providecommand{\BIBentryALTinterwordspacing}{\spaceskip=\fontdimen2\font plus
\BIBentryALTinterwordstretchfactor\fontdimen3\font minus
  \fontdimen4\font\relax}
\providecommand{\BIBforeignlanguage}[2]{{%
\expandafter\ifx\csname l@#1\endcsname\relax
\typeout{** WARNING: IEEEtran.bst: No hyphenation pattern has been}%
\typeout{** loaded for the language `#1'. Using the pattern for}%
\typeout{** the default language instead.}%
\else
\language=\csname l@#1\endcsname
\fi
#2}}
\providecommand{\BIBdecl}{\relax}
\BIBdecl

\bibitem{behrens2009diffusion}
T.~E. Behrens and H.~Johansen-Berg, \emph{Diffusion MRI: From quantitative
  measurement to in-vivo neuroanatomy}.\hskip 1em plus 0.5em minus 0.4em\relax
  Academic Press, 2009.

\bibitem{ning2016joint}
L.~Ning, K.~Setsompop, O.~Michailovich, N.~Makris, M.~E. Shenton, C.-F. Westin,
  and Y.~Rathi, ``A joint compressed-sensing and super-resolution approach for
  very high-resolution diffusion imaging,'' \emph{NeuroImage}, vol. 125, pp.
  386--400, 2016.

\bibitem{chen2018angular}
G.~Chen, B.~Dong, Y.~Zhang, W.~Lin, D.~Shen, and P.-T. Yap, ``Angular
  upsampling in infant diffusion mri using neighborhood matching in xq space,''
  \emph{Frontiers in Neuroinformatics}, vol.~12, p.~57, 2018.

\bibitem{cheng2015joint}
J.~Cheng, D.~Shen, P.~J. Basser, and P.-T. Yap, ``Joint {6D} kq space
  compressed sensing for accelerated high angular resolution diffusion {MRI},''
  in \emph{International Conference on Information Processing in Medical
  Imaging}.\hskip 1em plus 0.5em minus 0.4em\relax Springer, 2015, pp.
  782--793.

\bibitem{mani2015acceleration}
M.~Mani, M.~Jacob, A.~Guidon, V.~Magnotta, and J.~Zhong, ``Acceleration of high
  angular and spatial resolution diffusion imaging using compressed sensing
  with multichannel spiral data,'' \emph{Magnetic resonance in medicine},
  vol.~73, no.~1, pp. 126--138, 2015.

\bibitem{wu2019diffusion}
W.~Wu, P.~J. Koopmans, J.~L. Andersson, and K.~L. Miller, ``Diffusion
  acceleration with gaussian process estimated reconstruction ({DAGER}),''
  \emph{Magnetic resonance in medicine}, vol.~82, no.~1, pp. 107--125, 2019.

\bibitem{shi2016super}
F.~Shi, J.~Cheng, L.~Wang, P.-T. Yap, and D.~Shen, ``Super-resolution
  reconstruction of diffusion-weighted images using {4D} low-rank and total
  variation,'' in \emph{MICCAI Workshop on Computational Diffusion MRI}.\hskip
  1em plus 0.5em minus 0.4em\relax Springer, 2016, pp. 15--25.

\bibitem{tanno2017bayesian}
R.~Tanno, D.~E. Worrall, A.~Ghosh, E.~Kaden, S.~N. Sotiropoulos, A.~Criminisi,
  and D.~C. Alexander, ``Bayesian image quality transfer with {CNN}s: exploring
  uncertainty in d{MRI} super-resolution,'' in \emph{Medical Image Computing
  and Computer-Assisted Intervention (MICCAI)}.\hskip 1em plus 0.5em minus
  0.4em\relax Springer, 2017, pp. 611--619.

\bibitem{bruna2013spectral}
J.~Bruna, W.~Zaremba, A.~Szlam, and Y.~LeCun, ``Spectral networks and locally
  connected networks on graphs,'' \emph{arXiv preprint arXiv:1312.6203}, 2013.

\bibitem{henaff2015deep}
M.~Henaff, J.~Bruna, and Y.~LeCun, ``Deep convolutional networks on
  graph-structured data,'' \emph{arXiv preprint arXiv:1506.05163}, 2015.

\bibitem{goodfellow2014generative}
I.~Goodfellow, J.~Pouget-Abadie, M.~Mirza, B.~Xu, D.~Warde-Farley, S.~Ozair,
  A.~Courville, and Y.~Bengio, ``Generative adversarial nets,'' in
  \emph{Advances in Neural Information Processing Systems (NIPS)}, 2014, pp.
  2672--2680.

\bibitem{radford2015unsupervised}
A.~Radford, L.~Metz, and S.~Chintala, ``Unsupervised representation learning
  with deep convolutional generative adversarial networks,'' \emph{arXiv
  preprint arXiv:1511.06434}, 2015.

\bibitem{denton2015deep}
E.~L. Denton, S.~Chintala, R.~Fergus \emph{et~al.}, ``Deep generative image
  models using a laplacian pyramid of adversarial networks,'' in \emph{Advances
  in Neural Information Processing Systems (NIPS)}, 2015, pp. 1486--1494.

\bibitem{isola2017image}
P.~Isola, J.-Y. Zhu, T.~Zhou, and A.~A. Efros, ``Image-to-image translation
  with conditional adversarial networks,'' \emph{arXiv preprint}, 2017.

\bibitem{zhu2017unpaired}
J.-Y. Zhu, T.~Park, P.~Isola, and A.~A. Efros, ``Unpaired image-to-image
  translation using cycle-consistent adversarial networks,'' \emph{arXiv
  preprint}, 2017.

\bibitem{hong2019multifold}
Y.~Hong, G.~Chen, P.-T. Yap, and D.~Shen, ``Multifold acceleration of diffusion
  {MRI} via deep learning reconstruction from slice-undersampled data,'' in
  \emph{International Conference on Information Processing in Medical
  Imaging}.\hskip 1em plus 0.5em minus 0.4em\relax Springer, 2019, pp.
  530--541.

\bibitem{hong2019reconstructing}
------, ``Reconstructing high-quality diffusion {MRI} data from orthogonal
  slice-undersampled data using graph convolutional neural networks,'' in
  \emph{International Conference on Medical Image Computing and
  Computer-Assisted Intervention}.\hskip 1em plus 0.5em minus 0.4em\relax
  Springer, 2019, pp. 529--537.

\bibitem{defferrard2016convolutional}
M.~Defferrard, X.~Bresson, and P.~Vandergheynst, ``Convolutional neural
  networks on graphs with fast localized spectral filtering,'' in
  \emph{Advances in Neural Information Processing Systems (NIPS)}, 2016, pp.
  3844--3852.

\bibitem{bronstein2017geometric}
M.~M. Bronstein, J.~Bruna, Y.~LeCun, A.~Szlam, and P.~Vandergheynst,
  ``Geometric deep learning: going beyond {Euclidean} data,'' \emph{IEEE Signal
  Processing Magazine}, vol.~34, no.~4, pp. 18--42, 2017.

\bibitem{hammond2011wavelets}
D.~K. Hammond, P.~Vandergheynst, and R.~Gribonval, ``Wavelets on graphs via
  spectral graph theory,'' \emph{Applied and Computational Harmonic Analysis},
  vol.~30, no.~2, pp. 129--150, 2011.

\bibitem{chen2017neighborhood}
G.~Chen, B.~Dong, Y.~Zhang, D.~Shen, and P.-T. Yap, ``Neighborhood matching for
  curved domains with application to denoising in diffusion {MRI},'' in
  \emph{Medical Image Computing and Computer-Assisted Intervention
  (MICCAI)}.\hskip 1em plus 0.5em minus 0.4em\relax Springer, 2017, pp.
  629--637.

\bibitem{ronneberger2015u}
O.~Ronneberger, P.~Fischer, and T.~Brox, ``{U}-net: Convolutional networks for
  biomedical image segmentation,'' in \emph{International Conference on Medical
  image computing and computer-assisted intervention}.\hskip 1em plus 0.5em
  minus 0.4em\relax Springer, 2015, pp. 234--241.

\bibitem{dhillon2007weighted}
I.~S. Dhillon, Y.~Guan, and B.~Kulis, ``Weighted graph cuts without
  eigenvectors a multilevel approach,'' \emph{IEEE transactions on pattern
  analysis and machine intelligence}, vol.~29, no.~11, pp. 1944--1957, 2007.

\bibitem{he2016deep}
K.~He, X.~Zhang, S.~Ren, and J.~Sun, ``Deep residual learning for image
  recognition,'' in \emph{Computer Vision and Pattern Recognition (CVPR)},
  2016, pp. 770--778.

\bibitem{long2015fully}
J.~Long, E.~Shelhamer, and T.~Darrell, ``Fully convolutional networks for
  semantic segmentation,'' in \emph{Computer Vision and Pattern Recognition
  (CVPR)}, 2015, pp. 3431--3440.

\bibitem{nie20183}
D.~Nie, L.~Wang, E.~Adeli, C.~Lao, W.~Lin, and D.~Shen, ``{3-D} fully
  convolutional networks for multimodal isointense infant brain image
  segmentation,'' \emph{IEEE Transactions on Cybernetics}, 2018.

\bibitem{shi2016real}
W.~Shi, J.~Caballero, F.~Husz{\'a}r, J.~Totz, A.~P. Aitken, R.~Bishop,
  D.~Rueckert, and Z.~Wang, ``Real-time single image and video super-resolution
  using an efficient sub-pixel convolutional neural network,'' in
  \emph{Computer Vision and Pattern Recognition (CVPR)}, 2016, pp. 1874--1883.

\bibitem{chen2019prediction}
G.~Chen, Y.~Hong, K.~Huynh, W.~Lin, D.~Shen, and P.-T. Yap, ``Prediction of
  multi-shell diffusion {MRI} data using deep neural networks with diffusion
  loss,'' \emph{105th RSNA Scientific Assembly and Annual Meeting}, 2019.

\bibitem{tuch2004q}
D.~S. Tuch, ``Q-ball imaging,'' \emph{Magnetic Resonance in Medicine}, vol.~52,
  no.~6, pp. 1358--1372, 2004.

\bibitem{sotiropoulos2013advances}
S.~N. Sotiropoulos, S.~Jbabdi, J.~Xu, J.~L. Andersson, S.~Moeller, E.~J.
  Auerbach, M.~F. Glasser, M.~Hernandez, G.~Sapiro, M.~Jenkinson \emph{et~al.},
  ``Advances in diffusion {MRI} acquisition and processing in the human
  connectome project,'' \emph{NeuroImage}, vol.~80, pp. 125--143, 2013.

\bibitem{cciccek20163d}
{\"O}.~{\c{C}}i{\c{c}}ek, A.~Abdulkadir, S.~S. Lienkamp, T.~Brox, and
  O.~Ronneberger, ``{3D} {U}-{N}et: learning dense volumetric segmentation from
  sparse annotation,'' in \emph{Medical Image Computing and Computer-Assisted
  Intervention (MICCAI)}.\hskip 1em plus 0.5em minus 0.4em\relax Springer,
  2016, pp. 424--432.

\bibitem{zhang2012noddi}
H.~Zhang, T.~Schneider, C.~A. Wheeler-Kingshott, and D.~C. Alexander,
  ``{NODDI}: practical in vivo neurite orientation dispersion and density
  imaging of the human brain,'' \emph{Neuroimage}, vol.~61, no.~4, pp.
  1000--1016, 2012.

\bibitem{such2017robust}
F.~P. Such, S.~Sah, M.~A. Dominguez, S.~Pillai, C.~Zhang, A.~Michael, N.~D.
  Cahill, and R.~Ptucha, ``Robust spatial filtering with graph convolutional
  neural networks,'' \emph{IEEE Journal of Selected Topics in Signal
  Processing}, vol.~11, no.~6, pp. 884--896, 2017.

\end{thebibliography}

\end{document}